\documentclass[12pt,a4paper]{article}
\pdfoutput=1
\usepackage{jheppub}
\usepackage{bbm}
\usepackage{graphicx}
\usepackage{epsfig}
\usepackage[utf8]{inputenc}
\usepackage[T1]{fontenc}
\usepackage{amssymb,amsfonts,amsmath, mathtools}
\usepackage{tensor}
\usepackage{cases}
\usepackage{subfigure}

\newcommand{\bi}{\begin{itemize}}
\newcommand{\ei}{\end{itemize}}

\newcommand{\bea}{\begin{eqnarray}}
\newcommand{\eea}{\end{eqnarray}}
\newcommand{\be}{\begin{equation}}
\newcommand{\ee}{\end{equation}}
\newcommand{\ben}{\begin{eqnarray*}}
\newcommand{\een}{\end{eqnarray*}}
\newcommand{\bem}{\begin{pmatrix}}
\newcommand{\eem}{\end{pmatrix}}
\newcommand{\bl}{\begin{align}}
\newcommand{\el}{\end{align}}
\newcommand{\beg}{\begin{gather}}
\newcommand{\eeg}{\end{gather}}

\newcommand{\cB}{\mathcal{B}}

\newcommand{\cR}{\mathcal{R}}

\renewcommand{\a}{\alpha}
\renewcommand{\b}{\beta}
\renewcommand{\d}{\delta}

\newcommand{\m}{\mu}
\newcommand{\n}{\nu}

\renewcommand{\r}{\rho}                                     
\newcommand{\s}{\sigma}

\renewcommand{\O}{\Omega}

\newcommand{\pa}{\partial}

\newcommand{\Tr}{\mbox{Tr}}

\providecommand{\abs}[1]{\lvert#1\rvert}
\providecommand{\bd}[1]{\boldsymbol{#1}}

\title{\centerline{\Huge To $B$ or not to $B$:}
Primordial magnetic fields from Weyl anomaly}

\preprint{SISSA  32/2018/FISI}

\author[1, 2]{Andr\'e Benevides,}
\author[2, 3, 4]{Atish Dabholkar,}
\author[1, 2, 5]{and Takeshi Kobayashi}

\affiliation[1]{SISSA, Via Bonomea 265, 34136 Trieste, Italy}

\affiliation[2]{International Centre for Theoretical Physics\\
Strada Costiera 11, 34151 Trieste, Italy}

\affiliation[3]{Sorbonne Universit\'e, UPMC Univ Paris 06\\
  UMR 7589, LPTHE, Paris, F-75005 France}
  
\affiliation[4]{CNRS, UMR 7589, LPTHE,  Paris, F-75005 France}  

\affiliation[5]{INFN, Sezione di Trieste, Via Bonomea 265, 34136 Trieste, Italy}

\abstract{The quantum effective action for the electromagnetic field in an expanding universe has an anomalous dependence on the scale factor of the metric arising from virtual charged particles in the loops. It has been argued that this Weyl anomaly of quantum electrodynamics sources cosmological magnetic fields in the early universe. We examine this long-standing claim by using the effective action beyond the weak gravitational field limit which has recently been determined. We introduce a general criteria for assessing the quantumness of field fluctuations, and show that the Weyl anomaly is not able to convert vacuum fluctuations of the gauge field into classical fluctuations. We conclude that there is  \textit{no} production of coherent magnetic fields in the universe from the Weyl anomaly of quantum electrodynamics, irrespective  of the number of massless charged particles in the theory.} 

\begin{document}
\maketitle

\section{Introduction}

Cosmic inflation  freezes the quantum
fluctuations of the inflaton field into classical fluctuations which source the large-scale structures in the universe. 
While such a processing of field fluctuations happens generically both for nearly massless scalars and gravitons, 
the situation is different  for gauge fields. 
This is a simple consequence of the classical Weyl invariance of the 
Yang-Mills action. The dynamics of a gauge field governed by a Weyl invariant
action in a Friedmann-Robertson-Walker spacetime is independent of
the scale factor, and hence naively unaffected by the expansion of the universe.  

However, the classical Weyl invariance of the Yang-Mills action  is
violated in the quantum theory because of the need 
to regularize the path integral. These Weyl anomalies,  or equivalently
the nontrivial beta functions of the theory, imply that the quantum
effective action obtained after integrating out massless charged
particles is no longer Weyl invariant.  This is expected to lead  to an
anomalous dependence on the scale factor under a fairly mild assumption
that the masses of the charged particles that contribute to the quantum
loops  are negligible compared to the Hubble scale during the
cosmological era of interest.  For the Maxwell theory, the violation of Weyl invariance can lead to
gauge field excitations in the early universe, and thus to the generation
of electromagnetic fields.

In our universe, magnetic fields are observed on various scales such as
in galaxies and galaxy clusters.
Recent gamma ray observations suggest the presence of magnetic fields even in
intergalactic voids.
In order to explain the origin of the magnetic fields, theories of
primordial magnetogenesis 
have been studied in the literature, where most models
violate the Weyl invariance explicitly at the
classical level by coupling the gauge
field to some degrees of freedom beyond the Standard
Model of particle physics~\cite{Turner:1987bw, Ratra:1991bn}. 
See e.g. \cite{Kronberg:1993vk, Grasso:2000wj, Widrow:2002ud,
Barrow:2006ch, Kulsrud:2007an,Ryu:2011hu,Durrer:2013pga,
Subramanian:2015lua} for reviews on magnetic fields in the universe from different perspectives. 

It was pointed out in~\cite{Dolgov:1993vg} that 
the Weyl anomaly of quantum electrodynamics itself should also induce
magnetic field generation. 
If true, this would be a natural realization of primordial
magnetogenesis within the Standard Model.
Moreover, since the anomaly is intrinsic to the Standard Model, its
contribution to the magnetic fields, if any, is irreducible.
Hence it is important to evaluate this also for the purpose 
of identifying the minimum seed magnetic fields of our universe.
Since~\cite{Dolgov:1993vg}, there have indeed been many studies on this topic.
However, there is currently little consensus on the effect of the Weyl
anomaly on magnetic field generation.
One of the main difficulties in proceeding with these computations is
that the quantum effective action in curved spacetime is in general very
hard to evaluate.  In principle, it is a well-posed problem in perturbation theory.  One can regularize the path integral covariantly using dimensional regularization or short proper-time regularization and evaluate the effective action using the background field method. However, explicit evaluation of the path integral for a generic metric is not feasible.
For instance, to obtain the one-loop effective action it is necessary to compute the heat kernel of a Laplace-like operator in an arbitrary background, which amounts to solving the Schr\"odinger problem for an arbitrary potential.

One could  evaluate the effective action perturbatively in the weak field limit  using covariant nonlocal expansion of the heat kernel developed by Barvinsky, Vilkovisky, and collaborators \cite{Barvinsky:1984jd,Barvinsky:1985an}.  The effective action in this expansion has been worked out  to third order in curvatures \cite{Barvinsky:1988ds,Barvinsky:1994hw,Barvinsky:1994cg,Barvinsky:1995it}. 
Similar results have been obtained independently by Donoghue and
El-Menoufi \cite{Donoghue:2015xla,Donoghue:2015nba} using Feynman
diagrams.  Some of the earlier works on primordial magnetogenesis from
anomalies, e.g.~\cite{El-Menoufi:2015ztk}, relies
on the effective action derived in this weak field
approximation.
The weak field expansion is valid  in the regime $\cR^{2} \ll \nabla^{2} \cR  $, where $\cR$ denotes a \textit{generalized} curvature including both a typical geometric curvature $R$  as well as a typical gauge field strength $F$. During slow-roll inflation, one is in the regime of slowly varying geometric curvatures,  $R^{2} \gg \nabla^{2} R$, whereas during 
matter domination, one has $R^{2} \sim \nabla^{2} R$. Thus, during much
of the cosmological evolution, the curvatures are not weak compared to
their derivatives. Therefore, to study primordial magnetogenesis
reliably over a long range of cosmological evolution,  it is  essential
to overcome the limitations of the weak field
approximation.

It was shown recently in \cite{Bautista:2017enk}, that  one can go beyond the weak field approximation for Weyl flat spacetimes. In this case,  one can exploit Weyl anomalies and the symmetries of the background metric to  completely determine the  dependence of the effective action on the scale factor at one-loop even when the changes in the scale factor are large. The main advantage of this approach is that  Weyl anomalous dimensions of local operators  can be  computed reliably using  \textit{local} computations such as the Schwinger-DeWitt expansion \textit{without} requiring the weak field approximation $\cR^{2} \ll \nabla^{2} \cR $.   The resulting action obtained by integrating the anomaly is  necessarily nonlocal and essentially resums the Barvinsky-Vilkovisky expansion to all orders in curvatures albeit for the restricted class of Weyl-flat metrics. A practical advantage is that  one can extract  the essential physics with relative ease using only the local Schwinger-DeWitt expansion which is computationally much simpler. 

In this paper we use the quantum effective action of \cite{Bautista:2017enk} beyond the weak
field limit, and present the first consistent computation of the effect
of the Weyl anomaly on cosmological magnetic field generation.
We study $U(1)$~gauge fields originating as 
vacuum fluctuations in the inflationary universe, and analyze their
evolution during the inflation and post-inflation epochs. 
Our main conclusion is that there is {\it no} production of coherent
magnetic fields from the Weyl anomaly of quantum electrodynamics, contrary to the claims of previous works. 
Our results hold independently of the details of the
cosmological history, or of the number of massless charged particles in
the theory.
We  show, in particular,  that even if
there were extra charged particles in addition to those of the Standard
Model, the Weyl anomaly with an increased beta function still would not
produce any magnetic fields.

Since the time-dependence introduced by the Weyl anomaly is unusually weak,  the analysis of the 
(non)generation of magnetic fields requires careful consideration of the nature of the field
fluctuations, in particular whether they are classical or quantum. 
For this purpose, we  introduce general criteria
for assessing the quantumness of field fluctuations. 
Using these criteria, we find that the quantum
fluctuations of the gauge field actually do {\it not} get converted into
classical fluctuations.

The paper is organized as follows. In $\S\ref{Action}$ we review the derivation
\cite{Bautista:2017enk} of the one-loop quantum effective action for a Weyl-flat metric. 
In $\S\ref{sec:quant}$ we  canonically
quantize the gauge fields using this action and introduce the criteria for quantumness. 
In $\S\ref{sec:cosmo}$ we analyze the evolution of the gauge field in
the early universe and show that  there is no production of coherent magnetic fields. 
In $\S\ref{Discussion}$  we comment on the relation of our work to earlier works
and conclude with a discussion of possible extensions. 

\section{Nonlocal Effective Action for Quantum Electrodynamics\label{Action}}

In the early universe before the electroweak phase transition, quarks and leptons are massless\footnote{The expectation value of the Higgs field could fluctuate during inflation with an
amplitude of the order of the inflationary
Hubble scale. However, since most of the Yukawa couplings are small, the induced
masses for these fermions would still  be smaller than the Hubble scale
and thus could be treated as effectively massless.}. Consider the hypercharge   $U(1)$ gauge field  of the Standard Model coupled to these  massless Dirac fermions which we collectively denote by $\Psi$. The  classical Lorentzian action  in curved spacetime   is
\be\label{action-L}
S_0[g, A, \Psi] =  - \int   d^4x \, \sqrt{|g|} \, \left[\frac{1}{4\,e_0^2}  \, F_{\mu\nu} F^{\mu\nu} 
+  i \,\bar \Psi\, \Gamma^a\, e_a^\mu \,  D_\mu\Psi \right] \, ,
\ee
where $F_{\mu \nu} = \partial_\mu A_{\nu} - \partial_\nu A_{\mu}$, and $e_0^{2}$ is the bare charge. 
The covariant derivative is defined including both the gauge connection
$A_{\mu}$ and the spin connection in the spinor representation~$w_{\mu} ^{ab}$:
\be
D_{\mu}:=  \partial_{\mu} - \frac{i}{2} w_{\mu} ^{ab} J_{ab} - i Q A_{\mu} \,, 
\ee
where $\{J_{ab}\}$ are the Lorentz representation matrices and Q is
the quantized charge of the field in units of~$e_0$.   

Classically, this action is invariant under  Weyl transformation:
\be
g_{\m\n} \rightarrow e^{2 \xi (x)} g_{\m\n} \, , \quad
g^{\m\n} \rightarrow e^{-2 \xi (x)} g^{\m\n} \, , \quad
\Psi \rightarrow e^{-\frac{3}{2} \xi (x)} \Psi \, , \quad
A_{\mu} \rightarrow A_{\mu}\, .
\ee
The Weyl symmetry  is anomalous because in the quantum theory one must introduce a  mass scale $M$ to renormalize the theory which violates the Weyl invariance. The  Weyl anomaly introduces a coupling of the gauge field to the Weyl factor of the metric. To analyze its effects on the fluctuations one can proceed in two steps. One can  first perform the path integral over fermions treating  both the metric and the gauge field as  backgrounds.  The resulting  effective action for the electromagnetic field will include all quantum effects of fermions in loops. It is necessarily nonlocal because it is obtained by integrating out massless fields. One can then quantize  the gauge field using this effective action to study the propagation of photons including all vacuum polarization effects as well as interactions with the background metric. 

In flat spacetime, with $g_{\m\n}=\eta_{\m\n}$, the quantum effective
action can be computed using standard field theory methods. Up to  one loop order, the quadratic action for the gauge fields is given by\footnote{The quantum effective action in general contains higher powers of the field strength but the resulting nonlinearities will not be relevant for our purposes. }
\bea\label{flat-action}
S_{\mathrm{flat}}[\eta, A]  = -\frac{1}{4e^2} \int   d^{4}x  
 \left[  F_{\m\n}(x) F^{\m\n}(x) -  \tilde{\b} (e) \int d^{4}y
 \ F_{\m\n}(x) L(x-y) F^{\m\n}(y) \right] 
\eea
where $e^{2} \equiv e^{2}(M)$ is the coupling renormalized at a
renormalization scale $M$, and $\tilde{\beta} (e)$
is the beta function of $\log e$, i.e.,
\begin{equation}
  \frac{d \log e}{d \log M} = \tilde{\beta} (e).
\label{e-flow}
\end{equation}
The beta function of quantum electrodynamics takes positive values,
which is written as
\be\label{beta}
\tilde{\beta} (e) = \frac{b e^2}{2} ,
\quad \mathrm{where} \quad
b = \frac{\Tr \,  (Q^{2} ) }{6\pi^2} . 
\ee
Here the coefficient~$b$ is expressed
in terms of  the trace of the charge operator taken over all massless
charged fermions\footnote{The fluctuations of the photon field are related to the fluctuations of the hypercharge gauge field by a number of order unity that depends on the Weinberg angle. This distinction will not be important for our conclusions.}. 
To keep the discussion general, we will also allow for the possibility of  extra massless charged particles beyond the Standard
Model in the early universe, and treat the beta function as an arbitrary
positive parameter.  

The bilocal kernel in the second term of the action is defined by a Fourier transform:
\be
L(x-y) \equiv \langle x |\log \left(\frac{-\partial^{2}}{M^{2}}\right)
|y\rangle= \int \frac{d^{4 }p}{(2\pi)^{4}}e^{i p_{\mu} (x^{\mu}
-y^{\mu})}\, \log \left(\frac{p_{\nu} p^{\nu}}{M^{2}}\right)\, .
\ee
The action in position space may seem a bit unfamiliar but is more easily recognizable in momentum space where it takes the form
\be\label{flat-action-YM2}
S_{\mathrm{flat}}[\eta, A] =  - \frac{1 }{4\,e^{2}}\int
\frac{d^{4}p}{(2\pi)^4} \,  \eta^{\r\a} 
\eta^{\s\b} \, \, \,   \tilde F_{\r\s}(-p)  \, \left[1 - \tilde{\beta}
\log \left(\frac{p_{\nu} p^{\nu}}{M^{2}}\right) \right]  \, \tilde F_{\a\b}(p).
\ee 
In this form, one recognizes the first term as the classical action with renormalized coupling and the second term as the usual one-loop logarithmic running of the coupling constant.  

\begin{figure}[t] 
\begin{center}
\includegraphics[width=0.7\linewidth]{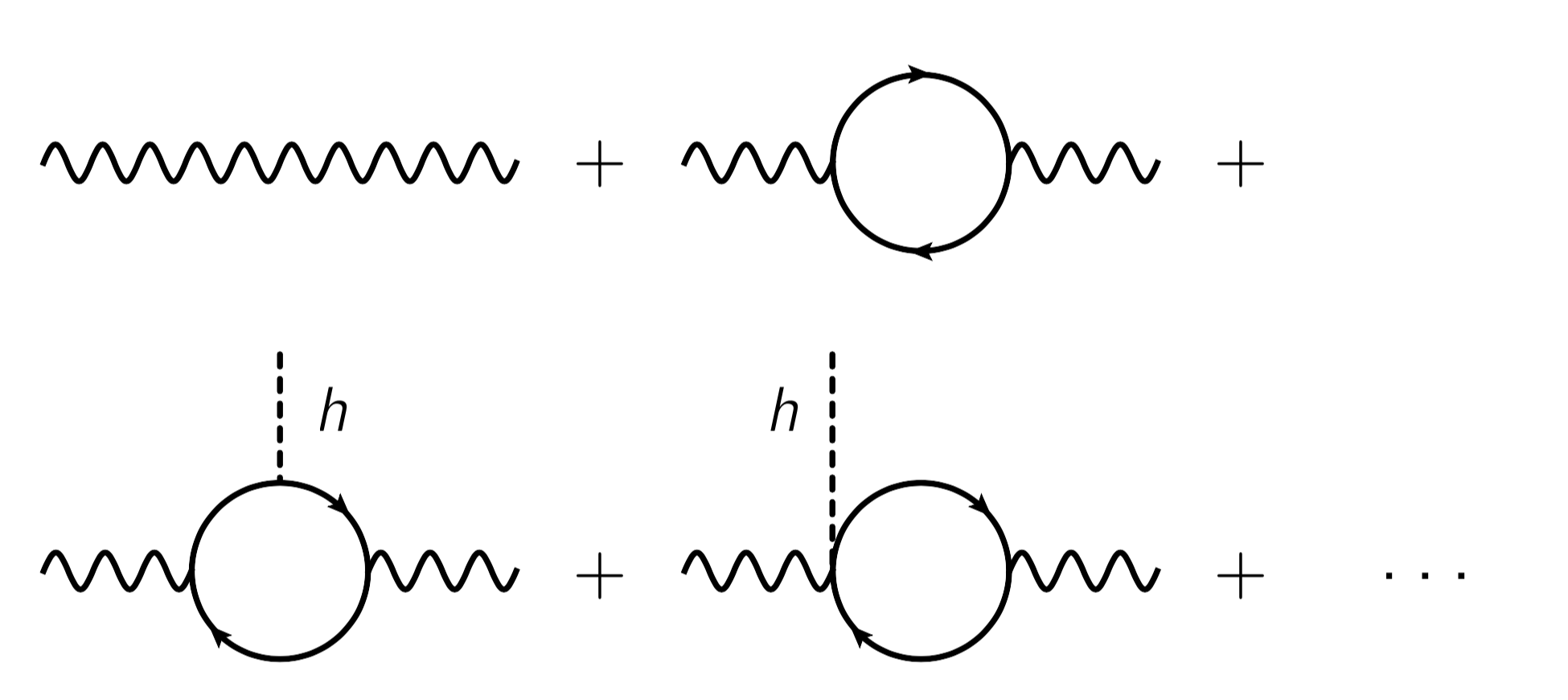}
\caption{The first term in the top line represents the classical propagation of the photon whereas the second term in the top line represents the one-loop correction to the propagator due to vacuum polarization in flat space. All diagrams in the bottom line  represent vacuum polarization in the presence of a curved metric $g_{\m\n} = \eta_{\m\n} + h_{\m\n}$ treating $h_{\m\n}$ as a perturbation. The Barvinsky-Vilkovisky expansion gives the covariantized nonlocal action resumming the specific powers of $h$ required for general covariance. Equation \eqref{anomaly-action} obtained by integrating the anomaly resums  these diagrams to all orders into a  simple expression for Weyl-flat spacetimes.}
\label{Prop}
\end{center}
\end{figure}

We are interested in the effective action in a curved spacetime. 
To get some intuition about the effect of the curvature, it is useful to consider the weak field limit so that the metric is close to being flat,  
$g_{\m\n} = \eta_{\m\n}  + h_{\m\n} $. If $h_{\m\n}$ is very small, then one can treat it  as a perturbation to  compute the corrections using Feynman diagrams. Various corrections arising from the interactions with the non-flat background metric are  shown diagrammatically in Figure~\ref{Prop} for the photon propagator. It is clear that even at one loop order, there are an infinite number of diagrams that contribute to the propagator. 
The Barvinsky-Vilkovisky expansion and related results complete the obtained expressions into non-linear and covariant functions of $h_{\m\n}$. Doing so however to a fixed order in $h_{\m\n}$ implies one neglects higher curvatures when compared to higher derivatives. More concretely,
\bea
R^2 \sim (\pa^2h)^2 , \quad
\nabla^2 R \sim  \pa^4 h.
\eea
As a result this `curvature expansion' is very different from the usual `derivative expansion' and  is  justified only in the limit $\nabla^2 R \gg R^2$.
If one is interested in a metric such as the Friedmann-Robertson-Walker metric that differs substantially from the Minkowski metric, a perturbative evaluation in this weak field limit clearly would not be adequate.

It was shown in  \cite{Bautista:2017enk}  that for Weyl-flat metrics, i.e., of the form 
$g_{\m\n}=e^{2\O}\eta_{\mu \nu}$,  it  is indeed possible  to
obtain the quantum effective action at one-loop as an \textit{exact}
functional of  $\Omega$ \textit{without} assuming small $h$. This is
achieved  by integrating the Weyl anomaly and matching with the flat
space results \cite{Bautista:2017enk}. The part of the action
that contains the gauge field takes a simple form\footnote{In the space of metrics, this action is evaluated in the subspace of Weyl-flat metrics.  For this reason it is beyond the reach of this method to compute the equations of motion for the background metric which requires a functional variation with respect to $g_{\m\n}$ even in directions orthogonal to the Weyl orbits.}:
\bea \label{ouraction}
S[g, A] = S_{\mathrm{flat}}[\eta,A] +  S_{\cB}[\eta, \O, A] \, ,
\eea
where $S_{\mathrm{flat}}$ is the effective action at one-loop as in
\eqref{flat-action}, and $S_{\cB}$ has the anomalous dependence on the
Weyl factor (or the scale factor in a Friedmann-Robertson-Walker spacetime):
\bea\label{anomaly-action}
S_{\cB}[\eta, \O, A]  =-\frac{\tilde{\beta}}{2e^{2}} \int d^4x\,
\O(x) \, F_{\m\n}(x) F^{\m\n}(x)  \,
\eea
where the indices are raised using the Minkowski metric as in \eqref{flat-action}. 
Thus the total effective action can be written as
\begin{equation}
 S = -\frac{1}{4e^2} \int d^4 x \, d^4 y\,  F_{\mu \nu}(x)\, 
  \langle x \rvert
  \left[ 1 - \tilde{\beta} 
   \log \left( \frac{-\partial^2}{M^2 \exp (2 \Omega(x)) } \right) \right]
  \lvert y \rangle\, 
  F^{\mu \nu} (y).
\label{action1}
\end{equation}

Even though the resummed answer of~(\ref{anomaly-action}) is a local functional of $\Omega$, it must come from nonlocal terms when expressed in terms of the original metric $g_{\m\n}$. There are non-local functionals that evaluate to the  Weyl factor $\Omega(x)$ on Weyl-flat backgrounds \cite{Fradkin:1978yf, Paneitz:2008, Riegert:1984kt}. One example is the Riegert functional:
\begin{equation}\label{Omega}
 \O[g](x)= \frac{1}{4}\int\! d^4 y\,\sqrt{|g(y)|}\, G_4(x,y)\, F_4[g](y) \, ,
\end{equation}
where 
\begin{align}\label{F4}
F_4[g] := E_4[g]-\frac{2}{3} \nabla^2 R[g]\,  = (R_{\m\n\r\s} R^{\m\n\r\s} - 4 R_{\m\n}R^{\m\n} +        R^{2}-\frac{2}{3} \nabla^{2} R )[g] \, ,
\end{align}
and the Green function $G_{4}(x,y)$ defined by 
\be
\Delta^{y}_4[g] G_{4}(x, y) =\frac{ \d^{(4)}(x -y) }{\sqrt{|g|}}\, 
\ee
is the inverse of the Weyl-covariant quartic differential operator 
\begin{equation}\label{diff-op}
 \Delta_4[g]= \left( \nabla^2 \right)^2 + 2 R^{\m\n}  \nabla_{\m}\nabla_{\n} + \frac{1}{3}\left(\nabla^{\n}R\right)\nabla_{\n} - \frac{2}{3}R\, \nabla^2 \, . 
\end{equation}
The expression~(\ref{Omega}) is manifestly covariant but nonlocal, consistent with
the fact that  the anomalous $\O$ dependence represents genuine
long-distance quantum effects that cannot be removed by counter-terms
that are local functionals of the metric. In the perturbative
Barvinsky-Vilkovisky regime we have  $ R^2 \ll \nabla^2 R $ and  one can expand the expression for $\O$ \eqref{Omega}  in curvatures to obtain to leading order 
\be {\O}[g] (x)= -\frac{1}{6}\frac{1}{\nabla^{2}}R + \dots \, . 
\label{Omegaweak}
\ee
It is clear from \eqref{Omega} that this expression receives corrections to all orders in $R$. The simple expression \eqref{ouraction} effectively resums these contributions to all orders as explained in  \cite{Bautista:2017enk}. 

\section{Quantization of the Gauge Field}
\label{sec:quant}

We now quantize the gauge field in a flat Friedmann-Robertson-Walker background,
\begin{equation}
 ds^2 = a(\tau)^2
  \left(-d \tau^2 + d \bd{x}^2 \right).
\end{equation}
Here the Weyl factor is $\Omega = \log a$, and thus the 
effective action~(\ref{action1}) takes the form
\begin{equation}
 S = -\frac{1}{4} \int d^4 x_1 d^4 x_2\,
  \mathcal{I}^2(x_1, x_2)
  F_{\mu \nu}(x_1)
  F^{\mu \nu}(x_2),
\label{action2}
\end{equation}
with
\begin{equation}
 \mathcal{I}^2(x_1, x_2) = \frac{1}{ e^2}
  \int \frac{d^4k}{(2 \pi )^4} e^{i k_\mu (x_1^\mu - x_2^\mu)} 
  \left[ 1  - \tilde{\beta} \log \left( \frac{a_{\star}^2}{a(\tau_1)^2} \right)
   - \tilde{\beta} \log \left( \frac{k_\nu k^\nu}{M^2 a_{\star}^2} \right) \right],
\label{Delta}
\end{equation}
where $x^\mu$ and $k^\mu$ are comoving coordinates and wave number,
respectively, and the indices are raised and lowered with the Minkowski metric.
We have introduced a scale factor~$a_{\star}$
and split the action into
$k_\nu k^\nu$-dependent and independent parts;
this splitting is completely arbitrary,
and hence $a_{\star}$ can also be chosen arbitrarily.

\subsection{Simplified Effective Action}

Let us decompose the spatial components of the gauge field into
irrotational and incompressible parts,
\begin{equation}
 A_\mu = (A_0, \partial_i S + V_i)
 \quad  \mathrm{with} \quad
  \partial_i V_i = 0,
\end{equation}
where we use Latin letters to denote spatial indices ($ i = 1,2,3$),
and the sum over repeated spatial indices is implied irrespective of
their positions. 
One can check that $A_0$ is a Lagrange multiplier, whose
constraint equation can be used to eliminate both $A_0$ and $S$ from the
action to yield
\begin{equation}
 S = \frac{1}{2} \int d^4 x_1 d^4 x_2\,
  \mathcal{I}^2(x_1, x_2)
  \left\{
   V_i'(x_1)  V_i'(x_2) -
   \partial_i V_j (x_1) \partial_i V_j (x_2)
  \right\},
\label{SofV}
\end{equation}
where we drop surface terms, 
and a prime denotes a derivative with respect to the conformal time~$\tau$.  
We now go to momentum space, 
\begin{equation}
 V_i (\tau, \boldsymbol{x}) = 
\sum_{p = 1,2}\int \frac{d^3 k}{(2 \pi )^3} \, 
e^{i \boldsymbol{k \cdot x}}\, 
\epsilon_i^{(p)} (\bd{k}) \, u_{\bd{k}}^{(p)} (\tau),
\label{fourier}
\end{equation}
where
$\epsilon_i^{(p)}(\boldsymbol{k})$ ($p = 1,2$) are 
two orthonormal polarization vectors that satisfy
\begin{equation}
 \epsilon_i^{(p)} (\boldsymbol{k}) \,  k_i = 0, 
\quad
 \epsilon_i^{(p)} (\boldsymbol{k}) \,  \epsilon_i^{(q)} (\boldsymbol{k})  =
 \delta_{pq}.
\label{2.10}
\end{equation}
From these conditions, it follows that
\begin{equation}
 \sum_{p = 1,2} \epsilon_i^{(p)} (\boldsymbol{k}) \, 
\epsilon_j^{(p)} (\boldsymbol{k})
 = \delta_{ij} - \frac{k_i k_j}{k^2},
\label{epepsum}
\end{equation}
where we use~$k$ to denote the amplitude of the
spatial wave number, i.e. $k = \abs{\bd{k}}$.
Unlike the spacetime indices, we do not assume implicit summation
over the polarization index~$(p)$.

The equation of motion of~$V_i$ requires the mode
function~$u_{\bd{k}}^{(p)} (\tau)$ to obey
\begin{equation}
\begin{split}
 0 = &
\left\{ 1 + 2 \tilde{\beta} \log \left( \frac{a(\tau )}{a_{\star}}
 \right) \right\} 
\left\{ u''^{(p)}_{\bd{k}}(\tau) + k^2 u^{(p)}_{\bd{k}}(\tau) \right\}
 + 2 \tilde{\beta} \frac{a'(\tau) }{a(\tau )}
u'^{(p)}_{\bd{k}}(\tau) \\
& -\tilde{\beta}
 \int d\tilde{\tau}
\left\{u''^{(p)}_{\bd{k}}(\tilde{\tau}) + k^2
 u^{(p)}_{\bd{k}}(\tilde{\tau})\right\}
\int \frac{dk^0}{2 \pi }e^{-i k^0 (\tau - \tilde{\tau} )}
\log \left( \frac{k_\mu k^\mu}{M^2 a_{\star}^2} \right).
\label{marui}
\end{split}
\end{equation}
In order to estimate the second line, let us make the crude assumption
that the $k^0$~integral amounts to the replacement
\begin{equation}
 \int \frac{dk^0}{2 \pi }e^{-i k^0 (\tau - \tilde{\tau} )}
  \log \left( \frac{k_\mu k^\mu}{M^2 a_{\star}^2} \right)
  \to \delta (\tau - \tilde{\tau}) 
\log \left( \frac{k^2}{M^2 a_{\star}^2} \right),
\label{approx}
\end{equation}
where the coefficient of $\delta (\tau - \tilde{\tau})$ is obtained
by integrating both sides over~$\tau$. 
Then comparing with the terms in the $\left\{ \, \right\}$ parentheses
in the first line of~(\ref{marui}),   
one sees that the second line is negligible when 
\begin{equation}
\left| 1 + 2 \tilde{\beta} \log \left( \frac{a}{a_{\star}}
 \right) \right|
\gg
 \left|
  \tilde{\beta}  \log \left( \frac{k^2}{M^2 a_{\star}^2} \right)
  \right|.
\label{kcond}
\end{equation}

The second line of the equation of motion follows from the 
$\log (k_\nu k^\nu)$ term of~(\ref{Delta}) in the action.
Hence as long as the wave modes of interest satisfy the
condition~(\ref{kcond}), we can ignore this term and use a
simplified effective action of 
\begin{equation}
 S_{\mathrm{simp}} = -\frac{1}{4} \int d^4 x\, 
 I(\tau)^2 \, 
  F_{\mu \nu}(x)  F^{\mu \nu} (x),
\label{Slocal}
\end{equation}
where
\begin{equation}
 I(\tau)^2 = \frac{1}{e^2}
\left[ 1 + 2 \tilde{\beta} \log
  \left(\frac{a(\tau)}{a_{\star}}\right)
\right].
  \label{I2}
\end{equation}
The equation of motion~(\ref{marui}) reduces to 
\begin{equation}
 u_{\bd{k}}''^{(p)} + 2 \frac{I'}{I} u_{\bd{k}}'^{(p)} + k^2 u_{\bd{k}}^{(p)} = 0.
\label{iv'}
\end{equation}
The action of the form~(\ref{Slocal}) with various time-dependent
functions~$I^2$ has been studied in the context of primordial
magnetogenesis since the seminal work of~\cite{Ratra:1991bn}.  
However we stress that, unlike many models of magnetogenesis 
whose time dependences are attributed to couplings to scalar fields
extraneous to the Standard Model,
here, the function~$I^2$ of~(\ref{I2}) arises from the Weyl anomaly of quantum
electrodynamics and thus is intrinsic to the Standard Model.
It should also be noted that, due to the positivity of the beta function
of quantum electrodynamics, $I^2$ monotonically increases in time. 

As is indicated by the equation of motion, there is no mixing between
different wave modes under the simplified action. 
This allows us to take the parameter~$a_{\star}$ differently for each wave
mode upon carrying out computations. A convenient choice we adopt is
\begin{equation}
 a_{\star} = \frac{k}{M},
  \label{akM}
\end{equation}
so that the simplifying condition~(\ref{kcond}) can be satisfied for a
sufficiently long period of time for every wave mode. 
However we should also remark that even with this choice,
the $\log (k_\nu k^\nu)$ term does not drop out completely.
This is because we have used the approximation~(\ref{approx}),
and thus in the right hand side of the
condition~(\ref{kcond}), the argument of the log should be 
considered to have some width around $k^2/M^2 a_{\star}^2$.
Hence we rewrite the simplifying condition for the choice of~(\ref{akM}),
by combining with the further assumption of $I^2 > 0$, as
\begin{equation}
 1 + 2 \tilde{\beta} \log \left( \frac{a M }{k} \right) \gg \tilde{\beta} .
\label{ii-iii'}
\end{equation}
If, on the other hand, the $\log (k_\nu k^\nu)$ term cannot be ignored,
this signals that the theory is strongly coupled\footnote{The
condition~(\ref{ii-iii'}) is rewritten as $I^2 \gg \tilde{\beta} / e^2 =
b/2$. Violating this condition provides an explicit example of
what is often referred to in the literature as the ``strong coupling 
problem'' of magnetogenesis with a tiny~$I$~\cite{Demozzi:2009fu}.}.
The Landau pole at which the coupling~$e$ blows up can be read off
from the running of the coupling~(\ref{e-flow}) as
\begin{equation}
 \Lambda_{\mathrm{max}} = M \exp\left( \frac{1}{2 \tilde{\beta}} \right).
\label{Landau}
\end{equation}
In terms of this, (\ref{ii-iii'}) is rewritten as
$k/a < \Lambda_{\mathrm{max}} \exp(-1/2)$.
Hence the simplifying condition can be understood as the requirement
that the physical momentum should be below the Landau pole 
during the times when one wishes to carry out computations.

The function~$\mathcal{I}^2$~(\ref{Delta}) in the full effective action
is independent of the renormalization scale~$M$,
since the coupling runs as~(\ref{e-flow}).
We note that with the choice~(\ref{akM}) for~$a_{\star}$,
the function~$I^2$~(\ref{I2}) in the simplified action also becomes
independent of~$M$.

\subsection{Canonical Quantization}

In order to quantize the gauge field, we promote~$V_i$ to an operator, 
\begin{equation}
 V_i(\tau, \boldsymbol{x}) = 
 \sum_{p = 1,2} \int \frac{d^3 k}{(2 \pi)^3} \, \epsilon^{(p)}_i (\boldsymbol{k})
\left\{
e^{i \boldsymbol{k \cdot x}}  a_{\boldsymbol{k}}^{(p)} 
u^{(p)}_{\boldsymbol{k}} (\tau) + 
e^{-i \boldsymbol{k \cdot x}} a_{\boldsymbol{k}}^{\dagger (p)}
u^{*(p)}_{\boldsymbol{k}} (\tau)  
\right\},
\label{Viop}
\end{equation}
where $a_{\boldsymbol{k}}^{(p)}$ and $a_{\boldsymbol{k}}^{\dagger (p)}$ 
are annihilation and creation operators satisfying the
commutation relations,
\begin{equation}
 [ a_{\boldsymbol{k}}^{(p)},\,  a_{\boldsymbol{l}}^{(q)} ] =
 [ a_{\boldsymbol{k}}^{\dagger (p)},\,  a_{\boldsymbol{l}}^{\dagger (q)}
 ] = 0,
\quad 
 [ a_{\boldsymbol{k}}^{(p)},\,  a_{\boldsymbol{l}}^{\dagger (q)} ] = (2
  \pi)^3 \, 
\delta^{pq} \,
\delta^{(3)}  (\boldsymbol{k} - \boldsymbol{l}) .
 \label{eq:commu}
\end{equation}
For $V_i$ and its conjugate
momentum which follows from the Lagrangian
$\mathcal{L} = (I^2/2) (V_i' V_i' - \partial_i V_j \partial_i V_j)$
(cf.~(\ref{SofV})) as
\begin{equation}
 \Pi_i = \frac{\partial \mathcal{L}}{\partial V_i'} = I^2 V_i',
\end{equation}
we further impose the commutation relations 
\begin{equation}\label{eq:commu3}
 \begin{split}
 &\left[ V_i(\tau, \boldsymbol{x}),\,  V_j (\tau, \boldsymbol{y}) \right] = 
 \left[ \Pi_i(\tau, \boldsymbol{x}),\,  \Pi_j (\tau, \boldsymbol{y}) \right]
 = 0,
\\
 &\left[ V_i(\tau, \boldsymbol{x}),\,  \Pi_j (\tau, \boldsymbol{y})
 \right] = i 
\delta^{(3)}   (\boldsymbol{x} - \boldsymbol{y})
\left( \delta_{ij} - \frac{\partial_i \partial_j}{\partial_l \partial_l} \right).
 \end{split}
\end{equation}
The second line can be rewritten using (\ref{epepsum}) as
\begin{equation}
 \left[ V_i(\tau, \boldsymbol{x}),\,  \Pi_j (\tau, \boldsymbol{y})
 \right] = 
 i \sum_{p = 1,2}\int \frac{d^3 k}{(2\pi)^3} \, 
 e^{i\boldsymbol{k\cdot}  (\boldsymbol{x - y})}
\epsilon_i^{(p)} (\boldsymbol{k}) \, \epsilon_j^{(p)} (\boldsymbol{k}).
\label{eq:commu4}
\end{equation}
Choosing the polarization vectors such that
\begin{equation}
 \epsilon^{(p)}_i
  (\boldsymbol{k}) = \epsilon^{(p)}_i (-\boldsymbol{k}),
  \label{poleps}
\end{equation}
one can check that the commutation relations~(\ref{eq:commu}) are
equivalent to (\ref{eq:commu3}) when the mode function 
is independent of the direction of~$\boldsymbol{k}$, i.e., 
\begin{equation}
 u_{\boldsymbol{k}}^{(p)} = u_k^{(p)},
\label{maru3}
\end{equation}
and also obeys
\begin{equation}
 I^2 \left(
u_k^{(p)} u'^{*(p)}_k - u_k^{*(p)} u'^{(p)}_k
\right) = i.
\label{v'}
\end{equation}
It follows from the equation of motion~(\ref{iv'}) that the left hand
side of this condition is time-independent, and thus this sets
the normalization of the mode function.

\subsection{Photon Number and Quantumness Measure}

Before proceeding to compute the cosmological evolution of the gauge
field fluctuations, we introduce two measures of  `quantumness' to determine when the field fluctuations  can  be regarded as classical. See
also~\cite{Grishchuk:1990bj,Green:2015fss} for discussions along similar lines.

In order to separately discuss each wave mode, we focus on the Fourier components
of the operator~$V_i$~(\ref{Viop}) and its conjugate momentum:
\begin{equation}
 \begin{split}
  V_i(\tau, \boldsymbol{x})
  &=   \sum_{p = 1,2} \int \frac{d^3 k}{(2 \pi)^3} \,
e^{i \boldsymbol{k \cdot x}}  \epsilon^{(p)}_i (\boldsymbol{k}) \, 
  v^{(p)}_{\boldsymbol{k}} (\tau),
  \\
  \Pi_i(\tau, \boldsymbol{x})
  &=   \sum_{p = 1,2} \int \frac{d^3 k}{(2 \pi)^3} \,
e^{i \boldsymbol{k \cdot x}}  \epsilon^{(p)}_i (\boldsymbol{k}) \, 
  \pi^{(p)}_{\boldsymbol{k}} (\tau) \, .
  \end{split}
\end{equation}
The Fourier modes  can be expressed in terms of the annihilation and creation operators as
\begin{equation}
  v^{(p)}_{\boldsymbol{k}} (\tau)  =  
a^{(p)}_{\boldsymbol{k}}  u^{(p)}_{k} (\tau) + 
a^{\dagger (p)}_{-\boldsymbol{k}}  u^{*(p)}_{k} (\tau),
\quad
  \pi^{(p)}_{\boldsymbol{k}} (\tau)  = 
I(\tau)^2 \left(
a^{(p)}_{\boldsymbol{k}}  u'^{(p)}_{k} (\tau) + 
a^{\dagger (p)}_{-\boldsymbol{k}}  u'^{*(p)}_{k} (\tau)
\right) .
\end{equation}
The commutation relations~(\ref{eq:commu}) or (\ref{eq:commu3}) entail
\begin{equation}
 [ v^{(p)}_{\boldsymbol{k}} (\tau),\,  v^{(q)}_{\boldsymbol{l}} (\tau) ]
  =
 [ \pi^{(p)}_{\boldsymbol{k}} (\tau),\,  \pi^{(q)}_{\boldsymbol{l}} (\tau) ] =
 0,
 \quad 
  [ v^{(p)}_{\boldsymbol{k}} (\tau),\,  \pi^{(q)}_{\boldsymbol{l}} (\tau) ] =
 i (2   \pi)^3 \, 
\delta^{pq} \,
\delta^{(3)}  (\boldsymbol{k} + \boldsymbol{l}) .
\end{equation}
We now introduce time-dependent annihilation and creation operators as
\begin{equation}
  b_{\bd{k}}^{(p)}(\tau)
  \equiv \sqrt{\frac{k}{2}} \, I(\tau) \, v^{(p)}_{\bd{k}}(\tau)
+ \frac{i}{\sqrt{2 k}} \frac{\pi^{(p)}_{\bd{k}}(\tau) }{I(\tau)},
\quad
  b_{\bd{k}}^{\dagger (p)}(\tau)
  \equiv \sqrt{\frac{k}{2}} \, I(\tau) \,  v^{(p)}_{-\bd{k}}(\tau)
- \frac{i}{\sqrt{2 k}} \frac{\pi^{(p)}_{-\bd{k}}(\tau) }{I(\tau)},
\end{equation}
so that $b_{\bd{k}}^{(p)}$ and $b_{\bd{k}}^{\dagger (p)}$ satisfy
equal-time commutation relations similar to (\ref{eq:commu}) of
$a_{\bd{k}}^{(p)}$ and $a_{\bd{k}}^{\dagger (p)}$, as well as
diagonalize the Hamiltonian,
\begin{equation}
 \tilde{H}  = \int d^3 x \left( \Pi_i V_i' - \mathcal{L} \right)
  = \sum_{p=1,2} \int \frac{d^3 k}{(2 \pi)^3} \, 
  k  \left(
b^{\dagger(p)}_{\bd{k}} b^{(p)}_{\bd{k}} 
+ \frac{1}{2}  [ b^{(p)}_{\bd{k}}, b^{\dagger (p)}_{\bd{k}}]
    \right). 
\end{equation}
The two sets of annihilation and creation operators are related by
\begin{equation}
  b_{\bd{k}}^{(p)}(\tau)
   = \alpha_{k}^{(p)} (\tau) \, a_{\bd{k}}^{(p)} +
   \beta_{k}^{*(p)} (\tau) \, a_{-\bd{k}}^{\dagger (p)}, 
\quad
  b_{\bd{k}}^{\dagger (p)}(\tau)
   = \alpha_{k}^{*(p)} (\tau) \, a_{\bd{k}}^{\dagger (p)} +
   \beta_{k}^{(p)} (\tau) \, a_{-\bd{k}}^{(p)}, 
\end{equation}
through time-dependent Bogoliubov coefficients:
\begin{equation}
 \alpha_k^{(p)} = I
  \left(\sqrt{\frac{k}{2}} \, u_k^{(p)} + \frac{i}{\sqrt{2 k}} \,
   u_k'^{(p)} \right),
  \quad
 \beta_k^{(p)} = I
  \left(\sqrt{\frac{k}{2}} \, u_k^{(p)} - \frac{i}{\sqrt{2 k}} \,
   u_k'^{(p)} \right).
\end{equation}
Using the normalization condition~(\ref{v'}), one can check that the
amplitudes of the coefficients obey
\begin{equation}
 \abs{\alpha_k^{(p)}}^2 - \abs{\beta_k^{(p)}}^2 = 1,
\label{a-minus-b}
\end{equation}
\begin{equation}
 \abs{\beta_k^{(p)}}^2 = \frac{I^2}{2}
  \left(
k \, \abs{ u_k^{(p)} }^2 +  \frac{\abs{ u_k'^{(p)} }^2}{k}
  \right) - \frac{1}{2}.
\label{beta-amp}
\end{equation}

When an adiabatic vacuum exists, 
$b^{\dagger(p)}_{\bd{k}} b^{(p)}_{\bd{k}}$ counts the numbers of photons
with polarization~$p$ and comoving momentum~$\bd{k}$.
However this operator itself is defined at all times, and it can be
interpreted as an instantaneous photon number.
Now let us suppose $a_{\bd{k}}^{(p)}$ and $a_{\bd{k}}^{\dagger (p)}$
to have initially diagonalized the Hamiltonian, i.e. $\beta_k^{(p)} =0$
in the distant past,
and that the system was initially in a vacuum state defined by
$ a_{\boldsymbol{k}}^{(p)} |0 \rangle = 0 $ for $p = 1,2$ and for all 
$\boldsymbol{k}$. Then at some later time, the number of
created photons per comoving volume is written as
\begin{equation}
 \frac{1}{V} \sum_{p = 1,2}\int \frac{d^3 k}{ (2 \pi)^3} \, 
  \langle 0| b^{\dagger(p)}_{\bd{k}} b^{(p)}_{\bd{k}} |0 \rangle
  = \sum_{p = 1,2} \int \frac{dk}{k}
\left(\frac{4 \pi }{3 k^3} \right)^{-1} \frac{2 }{3 \pi} \abs{\beta_k^{(p)}}^2,
\end{equation}
where $V \equiv \int d^3 x = (2 \pi)^3 \delta^{(3)} (\bd{0})$.
Thus one sees that $\frac{2 }{3 \pi} \abs{\beta_k^{(p)}}^2$ represents the
number of photons with polarization~$p$ and comoving
momentum of order\footnote{We assume that
$\abs{\beta_k^{(p)}}^2$ is smooth in~$k$ so that it does not have sharp
features in any narrow range of~$\Delta k \ll k$.}
$k$, within a comoving sphere of radius~$k^{-1}$.
The photon number~$\abs{\beta_k^{(p)}}^2 $  
(we will omit the coefficient~$\frac{2 }{3 \pi}$ as we are interested in
order-of-magnitude estimates)
is useful for judging whether magnetic field generation takes place:
A successful magnetogenesis model that gives rise to magnetic fields
with correlation length of~$k^{-1}$ does so by creating a large number
of photons with momentum~$k$, thus is characterized by
$\abs{\beta_k^{(p)}}^2  \gg 1$.  
On the other hand, if the photon number is as small as $ \abs{\beta_k^{(p)}}^2 =
\mathcal{O}(1)$, then it is clearly not enough to support coherent magnetic
fields in the universe. 

One can define another measure of quantumness introduced in~\cite{Green:2015fss}
(see also discussions in~\cite{Maldacena:2015bha}),
by the product of the standard deviations of $v_{\bd{k}}^{(p)}$ and
$\pi_{\bd{k}}^{(p)}$ in units of their commutator:
\begin{equation}
 \kappa_k^{(p)} (\tau)
  \equiv
 \left| \frac{
 \langle 0|
  v^{(p)}_{\boldsymbol{k}} (\tau)\, v^{(p)}_{-\boldsymbol{k}} (\tau)
 |0 \rangle\, 
 \langle 0|
  \pi^{(p)}_{\boldsymbol{k}} (\tau)\, \pi^{(p)}_{-\boldsymbol{k}} (\tau)
 |0 \rangle
 }{
  [ v^{(p)}_{\boldsymbol{k}}(\tau) , \, 
 \pi^{(p)}_{-\boldsymbol{k}} (\tau) ]^2
 }
 \right|^{1/2}
 =
  I(\tau)^2 \,  \left| u_k^{(p)} (\tau) \, u_k'^{(p)} (\tau) \right|.
\label{kappa}
\end{equation}
This quantity also corresponds to the classical volume of the space
spanned by $v_{\bd{k}}$ and 
$\pi_{-\bd{k}}$, divided by their quantum uncertainty\footnote{Here
$\kappa_k^{(p)}$ is defined slightly differently from the 
$\kappa$ introduced in Appendix~B.3
of~\cite{Green:2015fss}: $ \kappa = (2 \kappa_k^{(p)})^{-2}  $.}.
It takes a value of $\kappa_k^{(p)} \sim 1$
if the gauge field fluctuation with wave number~$k$ is quantum
mechanical.
On the other hand, if the fluctuations are effectively classical and large compared to the quantum uncertainty, then $\kappa_k^{(p)} \gg 1$.

This measure can also be used to quantify the conversion of 
quantum fluctuations into classical ones. 
As an example, consider a (nearly) massless scalar field in a de~Sitter
background (such as the inflaton), for which the measure~$\kappa_k$
can similarly be defined in terms of the scalar fluctuation and its
conjugate momentum.
Given that the fluctuation starts in a Bunch-Davies
vacuum when the wave mode~$k$ is deep inside the Hubble horizon,
one can check that $\kappa_k$ 
grows from $\sim 1$ when the wave mode is inside the horizon, to
$\kappa_k \gg 1 $ outside the horizon,
suggesting that the quantum fluctuations ``become classical'' upon
horizon exit.

The quantumness measure can also be expressed in terms of the Bogoliubov
coefficients as
\begin{equation}
(\kappa_k^{(p)} )^2 = \frac{1}{4}
  \left| (\alpha_k^{(p)})^2 - (\beta_k^{(p)})^2 \right|^2
=   \frac{1}{4} +
  \abs{\beta_k^{(p)}}^2
  \left( 1 +  \abs{\beta_k^{(p)}}^2 \right)
  \sin^2 \left\{ \mathrm{arg} (\alpha_k^{(p)} \beta_k^{*(p)} ) \right\},
\label{kappa-beta}
\end{equation}
where we have used~(\ref{a-minus-b}) upon moving to the far right hand side. 
This clearly shows that $\kappa_k^{(p)}$ takes its minimum value~$1/2$ 
when there is no photon production, i.e., for $\beta_k^{(p)} = 0$. 
Is is also useful to note that 
the instantaneous photon number~$\abs{\beta_k^{(p)}}^2$ corresponds to 
the sum of squares of the standard deviations of $v_{\bd{k}}^{(p)}$ and
$\pi_{\bd{k}}^{(p)}$ with weights~$(k I^2)^{\pm 1}$, cf.~(\ref{beta-amp}).
An inequality relation of
\begin{equation}
  \abs{\beta_k^{(p)}}^2 \geq \kappa_k^{(p)} - \frac{1}{2} 
\label{beta-kappa-ineq}
\end{equation}
is satisfied.

The classical Maxwell theory is described by setting $I^2 = 1/e^2$,
with which the mode function is a linear combination of plane waves.
Then $\abs{\beta_k^{(p)}}$ simply corresponds to the amplitude of the
coefficient of the negative frequency wave, and thus is time-independent.
It can also be checked in this case that 
$\mathrm{arg} (\alpha_k^{(p)} \beta_k^{*(p)} ) = -2 k \tau +
\mathrm{const.}$,
and hence one sees from~(\ref{kappa-beta}) that
$\kappa_k^{(p)}$ for plane waves oscillates in
time within the range $1/2 \leq \kappa_k^{(p)} \leq 1/2 +
\abs{\beta_k^{(p)}}^2$.

\section{Cosmological Evolution of the Gauge Field}
\label{sec:cosmo}

The anomalous dependence of the effective action for quantum electrodynamics
on the scale factor couples the gauge field to the cosmological expansion.
Here, in order to study the time evolution,
the initial condition of the gauge field needs to be specified.
A natural option is to start as quantum fluctuations
when the wave modes were once deep inside the Hubble horizon of the
inflationary universe.
This Bunch-Davies vacuum during the early stage of
the inflationary epoch will be the starting point of our computation.

It should also be noted that the scale factor dependence does not ``switch
off'', as long as there are massless particles around, and thus the cosmological background continues to affect the gauge
field equation of motion even after inflation.
(In this respect, the effect of the Weyl anomaly serves as a subclass of
the inflationary plus post-inflationary magnetogenesis scenario proposed
in~\cite{Kobayashi:2014sga}.)
After inflation ends, the universe typically enters an epoch dominated
by a harmonically oscillating inflaton field,
whose kinetic and potential energies averaged over the oscillation are
equal and thus behaves as pressureless matter. 
Then eventually the inflaton decays and heats up
the universe; during this reheating phase, the universe is expected
to become filled with charged particles and thus the gauge field evolution
can no longer be described by the source-free equation of
motion~(\ref{iv'}).
We also note that after the electroweak phase transition, the charged
particles in the Standard Model will obtain masses and therefore our effective
action~(\ref{ouraction}) becomes invalid. 
Hence the gauge field evolution will be followed up until the time of
reheating or electroweak phase transition, whichever happens earlier. 

\subsection{Bunch-Davies Vacuum}

For the purpose of obtaining a gauge field solution that corresponds to
the vacuum fluctuations,  
it is convenient to rewrite the equation of motion~(\ref{iv'}) into the
following form: 
\begin{equation}
 (I u_k)'' + \omega_k^2 \, I u_k = 0,
\quad \mathrm{where} \quad
 \omega_k = \left( k^2 - \frac{I''}{I} \right)^{1/2}.
\label{iv''}
\end{equation}
We have dropped the polarization index~$(p)$
because the action is symmetric between the two polarizations. 
This equation admits an approximate solution of the WKB-type
\begin{equation}
 u_k^{\mathrm{WKB}}(\tau) =
  \frac{1}{\sqrt{2 \omega_k (\tau)} \,  I (\tau)}
  \exp\left(
-i \int^\tau d\tilde{\tau}\,  \omega_k (\tilde{\tau})
      \right),
\label{WKB}
\end{equation}
given that the time-dependent frequency~$\omega_k$ satisfies the
adiabatic conditions,
\begin{equation}
 \left| \frac{\omega_k'}{\omega_k^2} \right|^2,
  \,
  \left| \frac{\omega_k''}{\omega_k^3} \right|
  \ll 1.
\label{vi}
\end{equation}
When further
\begin{equation}
 \omega_k^2 > 0,
  \label{vii}
\end{equation}
then $\omega_k$ is real and positive, and the WKB solution~(\ref{WKB}) 
describes a positive frequency solution
that satisfies the normalization condition~(\ref{v'}).
The period when the above conditions are satisfied can be understood by
noting that 
\begin{equation}
 \frac{I''}{k^2 I} =
  \frac{b }{ 2 I^2} \left( \frac{a H}{k} \right)^2
  \left( 1 + \frac{H'}{a H^2} - \frac{b }{2 I^2}\right).
\label{adi-exit}
\end{equation}
Here $H = a'/a^2$ is the Hubble rate.
The simplifying condition~(\ref{ii-iii'}) imposes $b \ll 2 I^2$,
and $|H' / a H^2| \lesssim 1$
is usually satisfied in a cosmological background.
Hence for wave modes that are inside the Hubble horizon,
i.e. $k > a H$,
it follows that $k^2 \gg |I'' / I|$.
This yields $\omega_k^2 \simeq k^2$,
satisfying the conditions (\ref{vi}) and (\ref{vii}).

In an inflationary universe, if one traces fluctuations with a fixed
comoving wave number~$k$ back in time, then its physical wavelength
becomes smaller than the Hubble radius.
Therefore we adopt the solution~(\ref{WKB})
when each wave mode was sub-horizon during inflation,
and take as the initial state the Bunch-Davies vacuum~$|0\rangle$
annihilated by~$a_{\bd{k}}$. 
Starting from this initial condition, we will see in the following
sections how the vacuum fluctuations evolve as the universe expands.

\subsection{Landau Pole Bound}

If we go back in time sufficiently far, the
physical momentum of a comoving mode~$k$ hits the Landau
pole~(\ref{Landau}) and we enter the strong coupling regime.
Here the simplifying condition~(\ref{ii-iii'}) also breaks down.
Hence in order to be able to set the Bunch-Davies initial condition
while maintaining perturbative control,
there needs to be a period during inflation when 
$k/a < \Lambda_{\mathrm{max}}$ as well as
the adiabaticity~(\ref{vi}) and stability~(\ref{vii}) conditions hold
simultaneously. 
We just saw that the conditions (\ref{vi}) and (\ref{vii}) hold
when the mode is sub-horizon, i.e. $k/a > H_{\mathrm{inf}}$,
where $H_{\mathrm{inf}}$ is the Hubble rate during
inflation. Therefore we infer a bound for the inflationary Hubble rate
\begin{equation}
 H_{\mathrm{inf}} < \Lambda_{\mathrm{max}},
  \label{HA}
\end{equation}
so that the Bunch-Davies vacuum can be adopted during the period of
$H_{\mathrm{inf}} < k/a < \Lambda_{\mathrm{max}}$.
We also see that this Landau pole bound on inflation collectively describes
the various conditions imposed in the previous sections, namely,
adiabaticity~(\ref{vi}) and stability~(\ref{vii}) during the
early stage of inflation, as well as the 
simplifying condition~(\ref{ii-iii'}) throughout the times of interest. 

The current observational limit on primordial gravitational waves sets
an upper bound on the inflation scale as 
$H_{\mathrm{inf}} \lesssim 10^{14}\, \mathrm{GeV}$~\cite{Akrami:2018odb}.
The Landau pole~$\Lambda_{\mathrm{max}}$ can be smaller than this
observational bound if there were sufficiently many massless charged
particles in the early universe. 
Taking for example the coupling to run through
$e^2 (M_Z) \approx 4 \pi/128$ at $ M_Z \approx 91.2\,
\mathrm{GeV}$~\cite{Patrignani:2016xqp},
a beta function coefficient as large as $b \gtrsim 0.4$ would lead to
$\Lambda_{\mathrm{max}} \lesssim 10^{14}\, \mathrm{GeV}$.
For such a large beta function, the adiabatic and perturbative regimes
cannot coexist for the gauge field during high-scale inflation.

\subsection{Slowly Running Coupling}

Before analyzing the gauge field evolution in full generality, 
let us first focus on cases with tiny beta functions.
Such cases can be treated analytically, by approximating
the $I^2$~function~(\ref{I2}) for small~$\tilde{\beta}$ as
\begin{equation}
 I^2 \simeq \frac{1}{e^2} \left( \frac{a}{a_{\star}} \right)^{2
  \tilde{\beta}}.
\label{Ik2power}
\end{equation}
We will later verify the validity of this approximation by comparing
with the results obtained from the original logarithmic~$I^2$.

In a flat Friedmann-Robertson-Walker universe with a constant equation
of state~$w$,
the equation of motion~(\ref{iv'}) under the power-law~$I^2$ 
admits solutions in terms of Hankel functions as~\cite{Kobayashi:2014sga},
\begin{equation}
 u_k = z^\nu \left\{
  c_1 H_{\nu}^{(1)} (z) + c_2 H_{\nu}^{(2)} (z)
		      \right\},
 \quad \mathrm{where} \quad
 z = \frac{2}{\abs{1 + 3w}} \frac{k}{a H} ,
\quad
 \nu = \frac{1}{2} - \frac{2 \tilde{\beta} }{1 + 3 w},
\label{uk-smallbeta}
\end{equation}
and the coefficients $c_1$, $c_2$ are independent of time.
Here, the equation of state parameter~$w$ can take any value except
for~$-1/3$, and the variable~$z$ scales with the scale factor as
$z \propto a^{(1+3 w)/2}$.
The time derivative of the mode function is written as
\begin{equation}
 u_k' = \mathrm{sign} (1+3w)\,  k z^\nu \left\{
  c_1 H_{\nu - 1}^{(1)} (z) + c_2 H_{\nu - 1}^{(2)} (z)
		      \right\}.
\label{ukprime-smallbeta}
\end{equation}
The behaviors of $u_k$ and $u_k'$ in the super-horizon limit,
i.e. $z \to 0$, can be read off from the asymptotic forms of the Hankel function:
\begin{equation} \label{Hanasym}
\begin{split}
 H_\nu^{(1)}(z) = \left( H_\nu^{(2)}(z)  \right)^* &\sim
  - \frac{i }{\pi } \,  \Gamma(\nu)
  \left(\frac{z}{2}\right)^{-\nu},
\\
 H_{\nu - 1}^{(1)}(z) = \left( H_{\nu - 1}^{(2)}(z) \right)^* &\sim
  - e^{(1 -\nu) \pi i } \, \frac{i}{\pi } \, \Gamma (1-\nu)
  \left(\frac{z}{2} \right)^{\nu-1},
\end{split}
\end{equation}
which are valid when $\tilde{\beta}$ is small such that $0 < \nu < 1$ is
satisfied.

\subsubsection*{During Inflation}

The inflationary epoch is characterized by the equation of state
$w = -1$ and a time-independent Hubble rate~$H_{\mathrm{inf}}$.
The solution that asymptotes to a positive frequency solution in the past is
\begin{equation}
 u_k =
\frac{1}{2 I} \left( \frac{\pi }{a H_{\mathrm{inf}}} \right)^{\frac{1}{2}}
\, H^{(1)}_{\frac{1}{2} + \tilde{\beta}} \left( \frac{k}{a H_{\mathrm{inf}}} \right),
    \label{uk_inf}
\end{equation}
whose normalization is set by (\ref{v'}) up to an unphysical phase.
Therefore the amplitudes of the mode function and its time derivative are
obtained as
\begin{equation}\label{ukasympinf}
\begin{split}
 k I^2 |u_k|^2 &= \frac{\pi k}{4 a H_{\mathrm{inf}}}
  \left| H^{(1)}_{ \frac{1}{2} + \tilde{\beta} } \left(\frac{k}{a
H_{\mathrm{inf}}}\right)
	  \right|^2
  \sim \frac{(\Gamma ( \frac{1}{2} + \tilde{\beta}) )^2}{2 \pi } 
  \left( \frac{2 a H_{\mathrm{inf}}}{k} \right)^{2 \tilde{\beta} },
\\
 \frac{1}{k} I^2 |u_k'|^2 &= \frac{\pi k}{4 a H_{\mathrm{inf}}}
  \left| H^{(1)}_{ - \frac{1}{2} + \tilde{\beta} } \left(\frac{k}{a
H_{\mathrm{inf}}}\right)
  \right|^2
  \sim \frac{(\Gamma (  \frac{1}{2} - \tilde{\beta} ) )^2}{2 \pi } 
  \left( \frac{2 a H_{\mathrm{inf}}}{k} \right)^{-2 \tilde{\beta} },
\end{split}
\end{equation}
where the far right hand sides show the asymptotic forms in the
super-horizon limit obtained by using~(\ref{Hanasym}). 
The geometric mean of these amplitudes yields the quantumness measure~(\ref{kappa}),
\begin{equation}
 \kappa_k = \frac{\pi k}{4 a H_{\mathrm{inf}}}
 \left|
 H^{(1)}_{ \frac{1}{2} + \tilde{\beta} } \left( \frac{k}{a H_{\mathrm{inf}}} \right) \,
 H^{(1)}_{ - \frac{1}{2} + \tilde{\beta}} \left( \frac{k}{a H_{\mathrm{inf}}} \right)
 \right|.
\end{equation}
In the sub-horizon limit $k/a H_{\mathrm{inf}} \to \infty$, this parameter
approaches $\kappa_k \sim 1/2 $ and thus the gauge field fluctuations are
quantum mechanical, which should be the case since we have started in
the Bunch-Davies vacuum.

The important question is whether the fluctuations become classical upon
horizon exit, as in the case for light scalar fields during inflation.
Using the reflection relation 
$\Gamma (\varpi ) \Gamma(1-\varpi) = \pi / \sin(\pi \varpi)$ for $\varpi
\notin \mathbb{Z}$, 
the asymptotic value of the quantumness parameter in the 
super-horizon limit $k/a H_{\mathrm{inf}} \to 0$ is obtained as
\begin{equation}
 \kappa_k \sim \frac{1}{2 \cos ( \pi \tilde{\beta} )}.
\label{kappa-cos2}
\end{equation}
Thus we find that $\kappa_k$ becomes time-independent outside the
horizon, and its asymptotic value depends\footnote{Since the approximate
expression~(\ref{Ik2power}) for~$I^2$
explicitly depends on the renormalization scale~$M$, so does the 
asymptotic value~(\ref{kappa-cos2}). However this $M$-dependence is tiny for slowly running  couplings.} only
on~$\tilde{\beta}$.
Most importantly, $\kappa_k$ is of order unity for $\tilde{\beta} \ll 1$.
This implies that if the beta function is small in the early universe, 
the time-dependence induced by the Weyl anomaly is not sufficient for converting 
vacuum fluctuations of the gauge field into classical ones.
Therefore no classical magnetic fields would arise.

We also estimate the instantaneous photon number~(\ref{beta-amp})
outside the horizon by summing the asymptotic expressions
of~(\ref{ukasympinf}), yielding
\begin{equation}
 \abs{\beta_k}^2 \sim 
\frac{(\Gamma ( \frac{1}{2} + \tilde{\beta} ) )^2}{4 \pi } 
  \left( \frac{2 a H_{\mathrm{inf}}}{k} \right)^{2 \tilde{\beta} }
+ \frac{(\Gamma ( \frac{1}{2} - \tilde{\beta}) )^2}{4 \pi } 
  \left( \frac{2 a H_{\mathrm{inf}}}{k} \right)^{-2 \tilde{\beta} }
- \frac{1}{2} .
\label{beta-super}
\end{equation}
The first term grows in time as $ \propto a^{2 \tilde{\beta}}$,
hence it will eventually dominate the right hand side if we wait long enough.
In a realistic cosmology, however, this term does not become much larger
than unity.
We will see this explicitly in the following sections.

\subsubsection*{After Inflation}
\label{subsec:after}

One can evaluate the mode function also in the effectively
matter-dominated epoch after inflation by matching solutions for $w =
-1$ and $w = 0$ at the end of inflation.
However let us take a simplified approach:
From the solutions (\ref{uk-smallbeta}) and (\ref{ukprime-smallbeta}) for generic~$w$,
and the asymptotic forms of the Hankel function~(\ref{Hanasym}),
one can infer the time-dependences of the mode function outside the
horizon in a generic cosmological background as
\begin{equation}
I^2 \abs{u_k}^2 \propto a^{2\tilde{\beta}}, \quad
 I^2 \abs{u_k'}^2 \propto a^{-2\tilde{\beta}}.
 \label{super-scaling}
\end{equation}
These super-horizon evolutions are determined only by the beta
function~$\tilde{\beta}$. Hence we find that for wave modes that exit the horizon
during inflation, the super-horizon expressions in~(\ref{ukasympinf})
continue to hold even after inflation, until the mode re-enters the
horizon.\footnote{This kind of argument
breaks down when the leading order approximations for the 
two Hankel functions~$H_{\nu}^{(1)}(z)$ and $H_{\nu}^{(2)}(z)$ 
cancel each other in the mode function.
Such cases are presented in~\cite{Kobayashi:2014sga}.
However in the current case where the power~$\tilde{\beta}$ of the
$I^2$~function is tiny, the cancellation does not happen as
we will see in the next section by comparing with numerical results
that the scaling~(\ref{super-scaling}) indeed holds until horizon re-entry.}
In particular, the super-horizon expressions (\ref{kappa-cos2})~for~$\kappa_k$
and (\ref{beta-super})~for~$\abs{\beta_k}^2$ also hold while the wave mode
is outside the horizon;
these expressions are the main results of the small-$\tilde{\beta}$ analysis.
Thus we find that $\kappa_k$ stays constant, while 
$\abs{\beta_k}^2$ basically continues to grow
until horizon re-entry.

If the mode re-enters the horizon before reheating and electroweak phase
transition, then we can continually use our effective action for
analyzing the gauge field dynamics.
Inside the horizon, where adiabaticity is recovered, 
the photon number~$\abs{\beta_k}^2$ becomes constant. 
On the other hand, the quantumness parameter~$\kappa_k$ oscillates in time
between $1/2$ and $ 1/2 + \abs{\beta_k}^2$, 
as described below~(\ref{beta-kappa-ineq}).

We see from~(\ref{beta-super}) that in order to have substantial photon
production, i.e. $\abs{\beta_k}^2 \gg 1$, the quantity~$(a H_{\mathrm{inf}} / k)^{2
\tilde{\beta}}$ needs to become large while the mode is outside the horizon. 
A larger~$H_{\mathrm{inf}}$ and $\tilde{\beta}$,
as well as a smaller~$k$ are favorable for this purpose. 
Here, for example, the magnitude of $a H_{\mathrm{inf}} / k$ upon the
electroweak phase transition at $T_{\mathrm{EW}} \sim 100\,
\mathrm{GeV}$ is, given that the universe has thermalized by
then, 
\begin{equation}
 \frac{a_{\mathrm{EW}} H_{\mathrm{inf}}}{k}
  \sim 10^{41} \left( \frac{H_{\mathrm{inf}}}{10^{14}\, \mathrm{GeV}} \right)
  \left( \frac{k}{a_0 } \cdot 10\, \mathrm{Gpc} \right)^{-1},
\label{aH-waru-k}
\end{equation}
where $a_0$ is the scale factor today.
The detailed value can be modified for different cosmological histories,
but what is relevant here is that even with the observably allowed highest
inflation scale~$H_{\mathrm{inf}} \sim 10^{14}\, \mathrm{GeV}$, and with
the size of the observable universe $a_0/k \sim 10\, \mathrm{Gpc}$
(or even on scales tens of orders of magnitude beyond that),
if the beta function is $\tilde{\beta} = \mathcal{O} (0.01)$,
then $(a_{\mathrm{EW}} H_{\mathrm{inf}} / k)^{2 \tilde{\beta}} =
\mathcal{O}(10)$.
Hence the number of photons created over the cosmological history 
would only be of $\abs{\beta_k}^2 =
\mathcal{O} (10)$, which is too small to support coherent magnetic fields.

On the other hand, the power-law~$I^2$~(\ref{Ik2power}) with
$\tilde{\beta} = 1$ yields an equation of motion equivalent to that of
a minimally coupled massless scalar field. 
Indeed, if one were to use the power-law~$I^2$ with $\tilde{\beta} \gtrsim 1$, 
then $\abs{\beta_k}^2$ and $\kappa_k$ 
are found to significantly grow outside the horizon; thus one would conclude
that gauge fluctuations do become classical and give rise to
cosmological magnetic fields for a large beta function.
However, in reality the power-law approximation breaks down when
$\tilde{\beta}$ is not tiny, and we will explicitly see in the next section
that the fluctuations of the gauge field actually never become
classical, independently of the value of~$\tilde{\beta}$.

\subsection{General Coupling}

In order to analyze quantum electrodynamics with generic beta functions, we have
numerically solved the equation of motion~(\ref{iv'}) 
for the original logarithmic $I^2$~function~(\ref{I2}),
with $a_{\star}$ chosen as~(\ref{akM}).
Starting from the WKB initial condition~(\ref{WKB})
during inflation when $H_{\mathrm{inf}} < k/a < \Lambda_{\mathrm{max}}$
is satisfied,
the mode function is computed in an inflationary as well as
the post-inflation matter-dominated backgrounds.
For the coupling we used
$e^2 (M_Z \approx 91.2\, \mathrm{GeV}) \approx 4 \pi/128$~\cite{Patrignani:2016xqp},
and considered it to run with a constant beta function
coefficient~$b$ in~(\ref{beta}).\footnote{In reality, $b$ is not a constant since the
number of effectively massless particles depends on the energy scale.
Moreover,  the hypercharge is related to the physical electric charge
through the Weinberg angle. However, these do not change the orders of
magnitude  of~$\beta_k$ and $\kappa_k$ for the electromagnetic field.}
For three light generations, $b$ is of order $0.1$.

\begin{figure}[htbp]
 \begin{center}
 \includegraphics[width=0.5\linewidth]{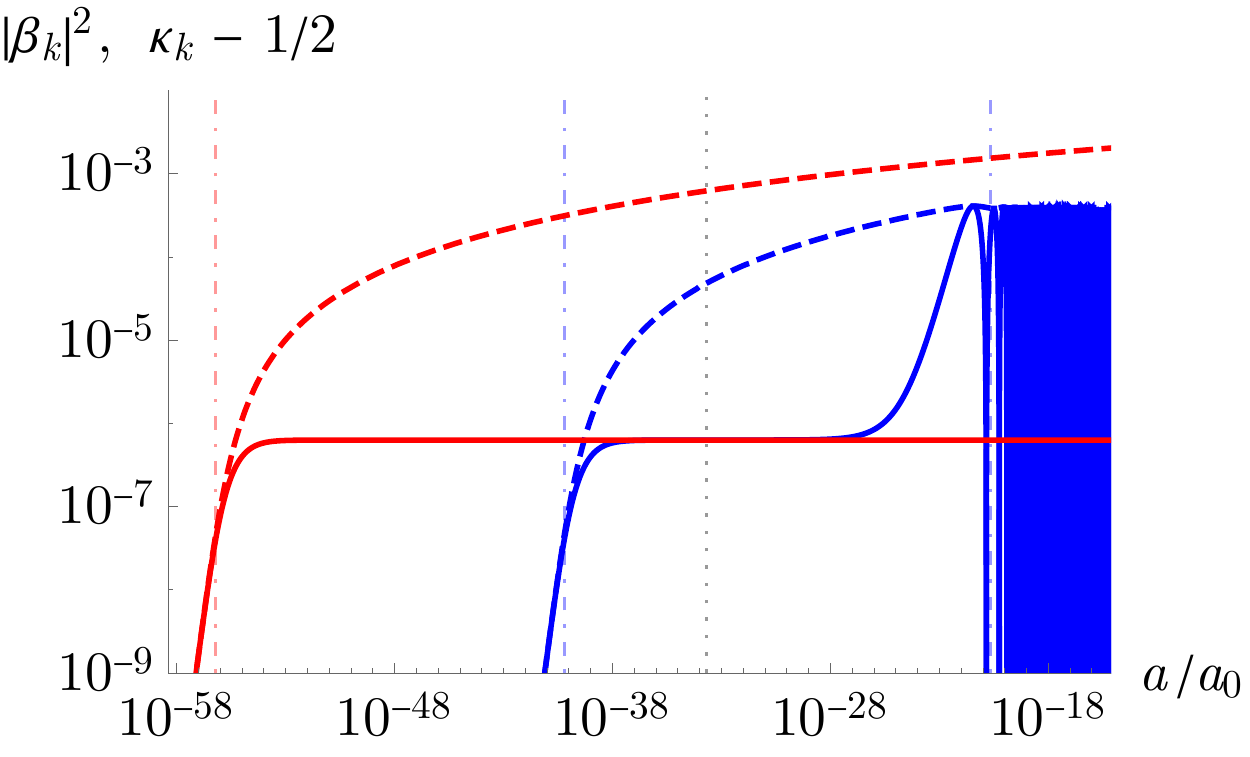}
  \caption{Time evolution of the instantaneous photon
  number~$\abs{\beta_k}^2$ (dashed lines) and quantumness
  parameter~$\kappa_k$ subtracted by~$1/2$ (solid lines),
  for wave numbers $k/a_0 = (10\, \mathrm{Gpc})^{-1}$ (red lines) and
  $(10^{-6}\, \mathrm{pc})^{-1}$ (blue lines).
  The beta function coefficient is set to $b = 0.01$.
  The background cosmology is fixed as $H_{\mathrm{inf}} = 10^{14}\,
  \mathrm{GeV}$ where inflation ends at the vertical dotted line,
  and reheating with $T_{\mathrm{reh}} = 100\, \mathrm{GeV}$
  taking place at the right edge of the plot.
  The wave mode $k/a_0 = (10\, \mathrm{Gpc})^{-1}$ exits the Hubble
  horizon at the vertical red dot-dashed line, whereas
  $k/a_0 = (10^{-6}\, \mathrm{pc})^{-1}$ exits and then re-enters the horizon
  at the blue dot-dashed lines.} 
 \label{fig:time-dep}
 \end{center}
\end{figure}

In Figure~\ref{fig:time-dep} we plot the evolution of 
$\abs{\beta_k}^2$ and $\kappa_k - 1/2$ as functions of the scale
factor~$a/a_0$.
Here the inflation scale is fixed to $H_{\mathrm{inf}} = 10^{14}\,
\mathrm{GeV}$,
and the reheating temperature to $T_{\mathrm{reh}}
= 100\, \mathrm{GeV}$ such that it coincides with the
scale of electroweak phase transition. 
The beta function coefficient is taken as $b = 0.01$
(thus $\tilde{\beta} (M_Z ) \approx 5 \times 10^{-4}$),
and the gauge field parameters are shown for two wave numbers:
$k/a_0 = (10\, \mathrm{Gpc})^{-1}$ (red lines) which corresponds to the size
of the observable universe today, and 
$k/a_0 = (10^{-6}\, \mathrm{pc})^{-1}$ (blue lines) which
re-enters the Hubble horizon before reheating. 
The figure displays the time evolution 
from when both modes are inside the horizon during inflation, until the
time of reheating. The vertical dotted line indicates the end of
inflation, and the dot-dashed lines for
the moments of Hubble horizon exit/re-entry.
With the beta function being tiny, the analytic expressions
(\ref{kappa-cos2}) and (\ref{beta-super}) derived in the 
previous section well describe the behaviors of
$\kappa_k$ and $\abs{\beta_k}^2$ outside the horizon.
After the mode $k/a_0 = (10^{-6}\, \mathrm{pc})^{-1}$
re-enters the horizon (after the blue dot-dashed line on the right), 
$\abs{\beta_k}^2$ becomes constant 
whereas $\kappa_k$ oscillates within the range of~(\ref{beta-kappa-ineq}). 
$\abs{\beta_k}^2$ is larger for smaller~$k$ as
there is more time for super-horizon evolution,
however even with $k/a_0 = (10\, \mathrm{Gpc})^{-1}$, 
$\abs{\beta_k}^2$ does not exceed unity by the time of reheating. 

\begin{figure}[htbp]
 \begin{minipage}{.48\linewidth}
  \begin{center}
 \includegraphics[width=\linewidth]{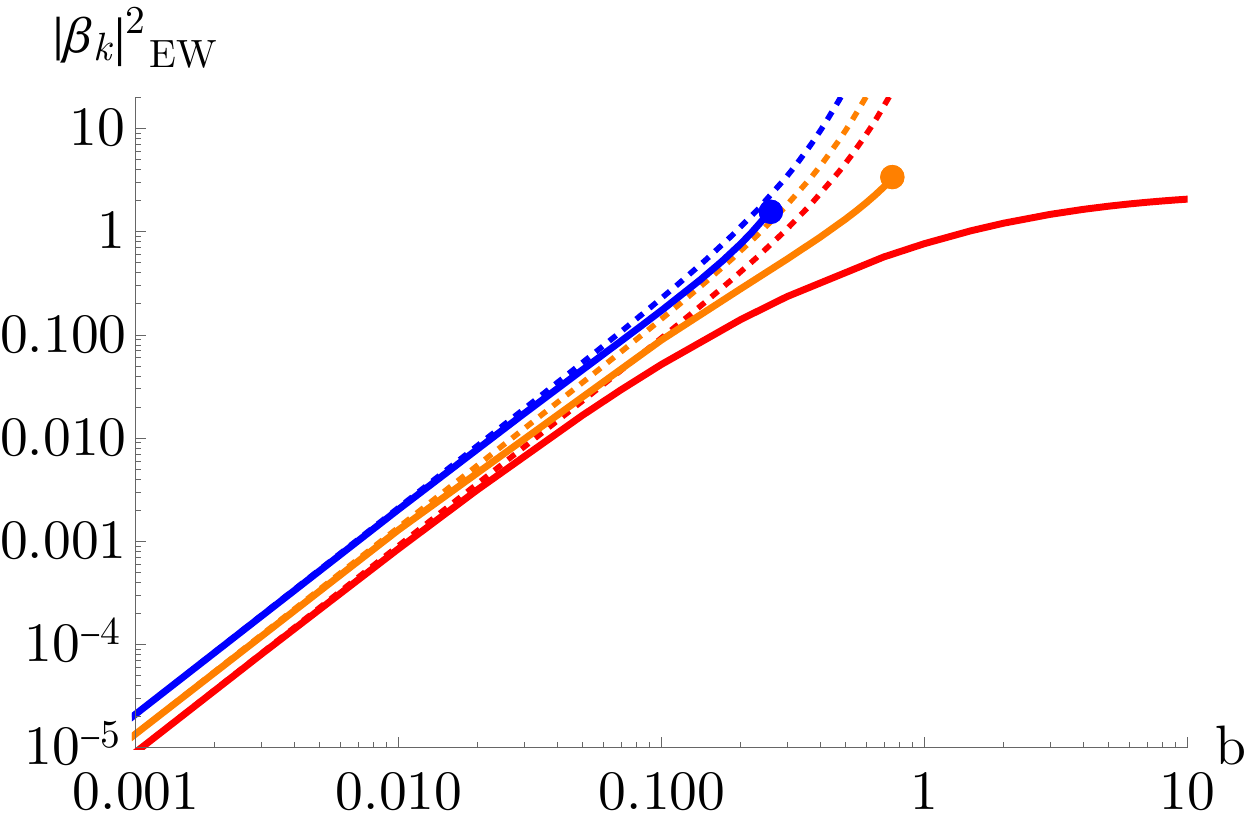}
  \end{center}
 \end{minipage} 
 \begin{minipage}{0.01\linewidth} 
  \begin{center}
  \end{center}
 \end{minipage} 
 \begin{minipage}{.48\linewidth}
  \begin{center}
 \includegraphics[width=\linewidth]{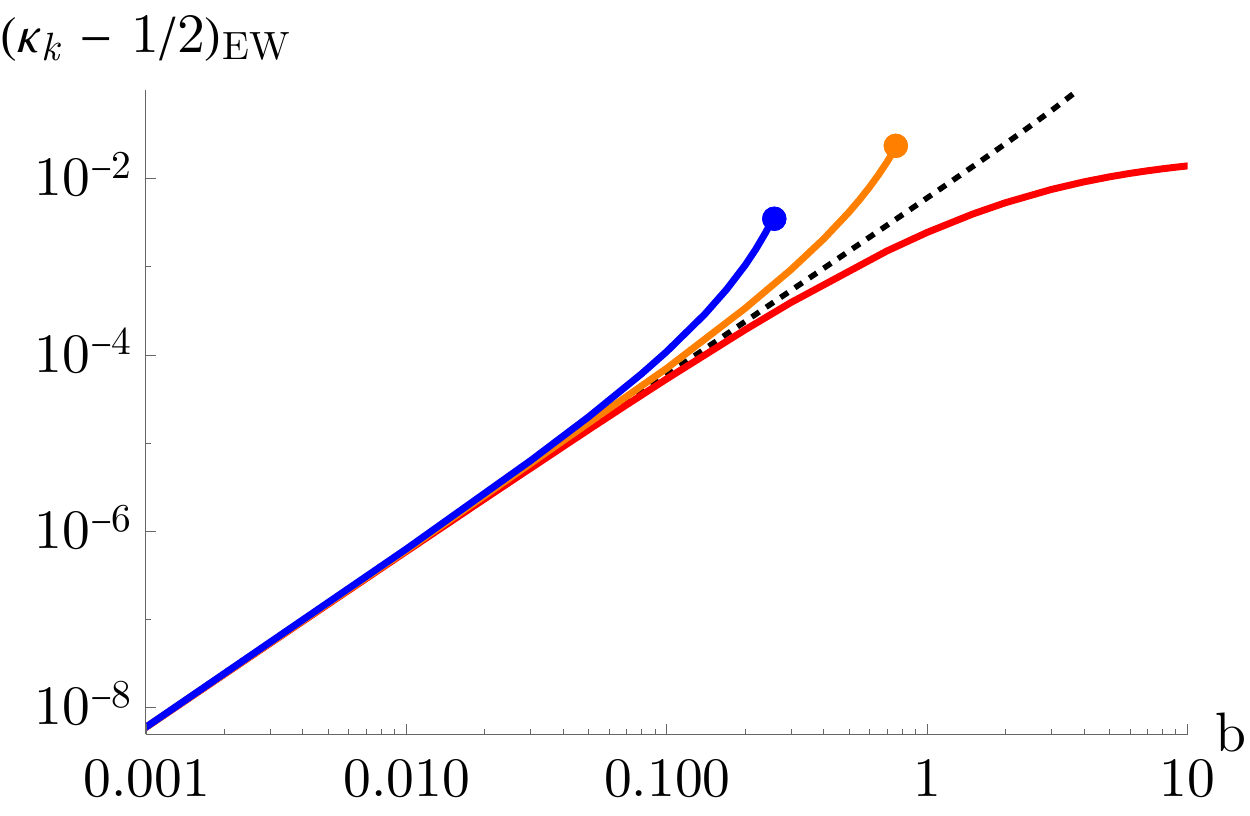}
  \end{center}
 \end{minipage} 
 \caption{Instantaneous photon number~$\abs{\beta_k}^2$ (left) and
 quantumness parameter~$\kappa_k$ subtracted by~$1/2$ (right) at the electroweak
 phase transition, as functions of the beta function coefficient~$b$. 
 The results are shown for a wave number $k/a_0 = (10\, \mathrm{Gpc})^{-1}$. 
 The reheating temperature is fixed to $T_{\mathrm{reh}} = 100\,
 \mathrm{GeV}$, while the inflation scale is varied as $H_{\mathrm{inf}}
 = 10^{14}\, \mathrm{GeV}$ (blue solid lines), $10^{6}\, \mathrm{GeV}$
 (orange solid), and $1\, \mathrm{GeV}$ (red solid). The endpoints of
 the curves show where the Landau pole bound is saturated (see text for
 details). The dashed lines show the analytic approximations derived for
 small beta functions: (\ref{beta-super})~for~$\abs{\beta_k}^2$, and
 (\ref{kappa-cos2})~for~$\kappa_k$.} 
 \label{fig:kappa}
\end{figure}

Figure~\ref{fig:kappa} shows 
$\abs{\beta_k}^2$ and $\kappa_k - 1/2$ as functions of the beta function
coefficient~$b$. Here the wave number is fixed to
$k/a_0 = (10\, \mathrm{Gpc})^{-1}$, and the reheating temperature to
$T_{\mathrm{reh}} = 100\,  \mathrm{GeV}$.
The gauge field parameters $\abs{\beta_k}^2$ and $\kappa_k$ in the
figure are evaluated at the electroweak phase transition, which coincides
with the time of reheating.
The solid curves with different colors correspond to different inflation
scales, which are 
chosen as $H_{\mathrm{inf}} = 10^{14}\, \mathrm{GeV}$ (blue),
$10^{6}\, \mathrm{GeV}$ (orange), and $1\, \mathrm{GeV}$ (red).
The Landau pole bound on the inflation scale~(\ref{HA}) imposes an upper
bound on the beta function coefficient as
$b_{\mathrm{max}} \approx 0.4$ for $H_{\mathrm{inf}} = 10^{14}\, \mathrm{GeV}$,
and $b_{\mathrm{max}} \approx 1.1$ for $H_{\mathrm{inf}} = 10^{6}\, \mathrm{GeV}$.
The computations have been performed for values of~$b$ 
up to $0.7 \times b_{\mathrm{max}}$,
which are shown as the endpoints of the blue and orange curves.
On the other hand, if the inflation scale is as low as
$H_{\mathrm{inf}} / 2 \pi \lesssim 100\, \mathrm{GeV}$,
the electroweak symmetry would already be broken during inflation.
However even in such cases, there might still be massless
charged particles in the early universe for some reason.
Hence for completeness, we have also carried out computations with
$H_{\mathrm{inf}} = 1\, \mathrm{GeV}$.
There is no Landau pole bound on~$b$ with such a low-scale inflation,
as is obvious from $H_{\mathrm{inf}}$ being smaller than
the scale~$M_Z$ where we set the coupling.
Hence this extreme case allows us to assess the implications of large
beta functions,
although it should also be noted that as one increases~$b$,
perturbation theory will eventually break down.

The wave mode $k/a_0 = (10\, \mathrm{Gpc})^{-1}$ for which the parameters
are evaluated is way outside the horizon at the electroweak phase
transition, thus the super-horizon approximations (\ref{kappa-cos2}) and
(\ref{beta-super}) should be valid for small beta functions.
These are shown as the dashed lines in the plots:
In the left panel, (\ref{beta-super}) is plotted using~(\ref{aH-waru-k}),
with the colors of the dashed lines corresponding to the different
inflation scales. In the right panel there is just one black dashed
line, because (\ref{kappa-cos2}) for~$\kappa_k$ only depends on the beta
function.\footnote{The expressions (\ref{kappa-cos2}) and (\ref{beta-super})
assume a tiny beta function, hence upon plotting the dashed lines,
the running is neglected and the coupling is fixed to $e^2 = 4 \pi/128$.}
The analytic approximations indeed agree well with the numerical results
at $b \lesssim 0.1$.
With larger~$b$, the numerical results
for $H_{\mathrm{inf}} = 10^{14}\, \mathrm{GeV}$ and $10^{6}\, \mathrm{GeV}$
show that even when approaching the Landau pole bound,
the parameters $\abs{\beta_k}^2 $ and $\kappa_k $ are at most of order
unity.
For $H_{\mathrm{inf}} = 1\, \mathrm{GeV}$
(assuming the existence of massless charged particles),
$\abs{\beta_k}^2 $ and $\kappa_k $ become less sensitive to~$b$ at $b
\gtrsim 1$ and thus turns out not to exceed order unity even with large~$b$.
Here we have focused on a rather
small wave number $k/a_0 = (10\, \mathrm{Gpc})^{-1}$
and a low reheating temperature $T_{\mathrm{reh}} = 100\, \mathrm{GeV}$; 
however for larger $k$ and $T_{\mathrm{reh}}$, 
the value of $\abs{\beta_k}^2 $ upon reheating becomes
even smaller as there is less time for the super-horizon evolution.

A heuristic argument for why the logarithmic $I^2$~function~(\ref{I2})
never leads to substantial photon production goes as follows:
Even if the beta function is as large as $\tilde{\beta} = 1$,
while the universe expands by, say, 100~$e$-foldings starting from~$a_\star$, 
the logarithmic $I^2$ grows only by a factor of~$200$.
On the other hand, if one were to obtain the same growth rate with a
power-law~$I^2$~(\ref{Ik2power}), 
the power would have to be as small as
$\tilde{\beta} \approx 0.03$;
then it is clear from the expressions (\ref{kappa-cos2}) and
(\ref{beta-super}) that the effect on photon production is tiny.

Thus we find that, for generic values of the beta function, 
inflation/reheating scales, and wave number,
the instantaneous photon number~$\abs{\beta_k}^2 $ and quantumness
measure~$\kappa_k$ do not become much greater than unity. 
Here, the physical meaning of~$\abs{\beta_k}^2 $ may seem ambiguous when the
wave mode is outside the horizon and thus an adiabatic vacuum is absent.
However as was discussed above~(\ref{aH-waru-k}),
the quantity~$\abs{\beta_k}^2 $ needs to become large while outside the horizon
in order to have a large number of photons to support
coherent magnetic fields. Moreover, the quantumness measure~$\kappa_k $
is bounded from above by~$\abs{\beta_k}^2 + 1/2 $, cf.~(\ref{beta-kappa-ineq}). 
Therefore we can conclude that,
unless some additional process significantly excites the gauge field after the
electroweak phase transition or reheating (namely, after our effective
action becomes invalid), 
the Weyl anomaly does not convert vacuum
fluctuations of the gauge field into classical fluctuations, let alone coherent
magnetic fields in the universe.

\section{Conclusions and Discussion \label{Discussion}}

We have analyzed cosmological excitation of magnetic fields due to the
Weyl anomaly of quantum electrodynamics.
Despite the anomalous dependence of the quantum effective action
on the scale factor of the metric,
we showed that the vacuum fluctuations of the gauge field do not get
converted into classical fluctuations,
as long as inflation happens at scales below the Landau pole.
In particular, the number of photons with a comoving momentum~$k$
produced within a comoving volume~$k^{-3}$ was found to be at most
of order unity, for generic~$k$. 
With such a small number of created photons,
we conclude that the Weyl anomaly does not give rise to
coherent magnetic fields in the universe. 
Our conclusion is independent of the details of the cosmological
history, or the number of massless charged particles in the theory.

For obtaining this result, which disproves the claims of many
previous works, there were two key ingredients.
The first was the quantum effective
action beyond the weak gravitational field limit.
We saw that, especially for cases where the beta function of quantum
electrodynamics was large in the early universe,
one could draw dramatically incorrect conclusions from
inappropriate assumptions about the effective action.
The essential point is that the anomalous dependence of the effective
action on the metric is associated to the renormalization group flow of
the gauge coupling, 
and therefore the dependence is only logarithmic in the scale factor,
cf.~(\ref{action2}) and (\ref{Delta});
this is in contrast with the case of massless scalar fields having
power-law dependences on the scale factor at the classical level.
The second element was a proper  evaluation of the nature  of the gauge field fluctuations, which we  discussed quantitatively in terms of the
photon number~(\ref{beta-amp}) and the quantumness parameter~(\ref{kappa}). 
Focusing on these quantities, we explicitly showed that the
logarithmic dependence on the background metric induced by the Weyl
anomaly does not lead to any generation of coherent classical magnetic fields. 

We now briefly comment on some of the earlier works on Weyl anomaly-driven
magnetogenesis. 
The original works~\cite{Dolgov:1981nw,Dolgov:1993vg} approximated the effect from the
Weyl anomaly as a power-law~$I^2$ for a generic beta function,
and thus arrived at the incorrect conclusion that a large beta function
gives rise to observably large magnetic fields.
On the other hand, the recent work~\cite{El-Menoufi:2015ztk} relies on
the effective action derived in the weak gravitational field limit.
The Weyl factor in an inflationary background is computed using the
curvature expansion of~(\ref{Omegaweak}), which yields
$\Omega \sim (2/3) \log a$ in the asymptotic
future, instead of the exact answer of $\log a$. 
At any rate, a logarithmic~$I^2$ is obtained with a form similar
to~(\ref{I2}) up to numerical coefficients.
However, the fact that a logarithmic~$I^2$ cannot produce enough photons
to support coherent magnetic fields was overlooked. 

Our considerations can also be applied to quantum chromodynamics. The
effective action is analogous to \eqref{ouraction} with $\tilde{\beta}$
given by the beta function of quantum chromodynamics coupled to massless
quarks.
One main difference from electrodynamics is that the beta function is
negative, yielding asymptotic freedom;
hence the theory goes into the
strongly coupled regime in the late universe.
The time evolution of the mode function can further be altered by 
the nonlinearities of the Yang-Mills action.
Here, since the dependence of the effective action on the scale factor is
anyway logarithmic, it may turn out that color magnetic fields are also
not generated by the Weyl anomaly;
however, it would be worthwhile to analyze systematically the range of
possibilities that can arise for $SU(N)$ Yang-Mills fields. 
With such analyses, one should also be able to evaluate the effect of the possible
mixing of the $SU(2)$ gauge field fluctuations into the photons upon the
electroweak phase transition, which we did not consider in this paper.
The study of  the
effects of the Weyl anomaly in the strongly coupled regime, 
for instance electrodynamics with inflation scales higher than the
Landau pole (thus with a very large beta function),
or chromodynamics near  the confinement transition is very interesting but would require nonperturbative methods. 

Even though the Weyl anomaly does not generate
coherent magnetic fields in the universe, it can produce a small
number of photons in the squeezed state. 
The squeezed light from the Weyl anomaly may have interesting consequences for
astrophysical observations~\cite{Grishchuk:1990bj,Allen:1999xw}.
Our criteria for quantumness could also be useful for studying  field excitations in other processes with weak time dependence.

\acknowledgments
We thank Teresa Bautista, Lorenzo Bordin,  Paolo Creminelli, Eiichiro Komatsu, Chunshan
Lin, and Giovanni Villadoro for helpful discussions.
T.K. acknowledges support from the INFN INDARK PD51 grant.


\bibliographystyle{JHEP}
\bibliography{weyl}

\providecommand{\href}[2]{#2}\begingroup\raggedright\begin{thebibliography}{10}

\bibitem{Turner:1987bw}
M.~S. Turner and L.~M. Widrow, {\it {Inflation Produced, Large Scale Magnetic
  Fields}},  {\em Phys. Rev.} {\bf D37} (1988) 2743.

\bibitem{Ratra:1991bn}
B.~Ratra, {\it {Cosmological 'seed' magnetic field from inflation}},  {\em
  Astrophys. J.} {\bf 391} (1992) L1--L4.

\bibitem{Kronberg:1993vk}
P.~P. Kronberg, {\it {Extragalactic magnetic fields}},  {\em Rept. Prog. Phys.}
  {\bf 57} (1994) 325--382.

\bibitem{Grasso:2000wj}
D.~Grasso and H.~R. Rubinstein, {\it {Magnetic fields in the early universe}},
  {\em Phys. Rept.} {\bf 348} (2001) 163--266,
  [\href{http://arxiv.org/abs/astro-ph/0009061}{{\tt astro-ph/0009061}}].

\bibitem{Widrow:2002ud}
L.~M. Widrow, {\it {Origin of galactic and extragalactic magnetic fields}},
  {\em Rev. Mod. Phys.} {\bf 74} (2002) 775--823,
  [\href{http://arxiv.org/abs/astro-ph/0207240}{{\tt astro-ph/0207240}}].

\bibitem{Barrow:2006ch}
J.~D. Barrow, R.~Maartens, and C.~G. Tsagas, {\it {Cosmology with inhomogeneous
  magnetic fields}},  {\em Phys. Rept.} {\bf 449} (2007) 131--171,
  [\href{http://arxiv.org/abs/astro-ph/0611537}{{\tt astro-ph/0611537}}].

\bibitem{Kulsrud:2007an}
R.~M. Kulsrud and E.~G. Zweibel, {\it {The Origin of Astrophysical Magnetic
  Fields}},  {\em Rept. Prog. Phys.} {\bf 71} (2008) 0046091,
  [\href{http://arxiv.org/abs/0707.2783}{{\tt arXiv:0707.2783}}].

\bibitem{Ryu:2011hu}
D.~Ryu, D.~R.~G. Schleicher, R.~A. Treumann, C.~G. Tsagas, and L.~M. Widrow,
  {\it {Magnetic fields in the Large-Scale Structure of the Universe}},  {\em
  Space Sci. Rev.} {\bf 166} (2012) 1--35,
  [\href{http://arxiv.org/abs/1109.4055}{{\tt arXiv:1109.4055}}].

\bibitem{Durrer:2013pga}
R.~Durrer and A.~Neronov, {\it {Cosmological Magnetic Fields: Their Generation,
  Evolution and Observation}},  {\em Astron. Astrophys. Rev.} {\bf 21} (2013)
  62, [\href{http://arxiv.org/abs/1303.7121}{{\tt arXiv:1303.7121}}].

\bibitem{Subramanian:2015lua}
K.~Subramanian, {\it {The origin, evolution and signatures of primordial
  magnetic fields}},  {\em Rept. Prog. Phys.} {\bf 79} (2016), no.~7 076901,
  [\href{http://arxiv.org/abs/1504.02311}{{\tt arXiv:1504.02311}}].

\bibitem{Dolgov:1993vg}
A.~Dolgov, {\it {Breaking of conformal invariance and electromagnetic field
  generation in the universe}},  {\em Phys. Rev.} {\bf D48} (1993) 2499--2501,
  [\href{http://arxiv.org/abs/hep-ph/9301280}{{\tt hep-ph/9301280}}].

\bibitem{Barvinsky:1984jd}
A.~O. Barvinsky and G.~A. Vilkovisky, {\it {The Generalized Schwinger-Dewitt
  Technique and the Unique Effective Action in Quantum Gravity}},  {\em Phys.
  Lett.} {\bf B131} (1983) 313--318.

\bibitem{Barvinsky:1985an}
A.~O. Barvinsky and G.~A. Vilkovisky, {\it {The Generalized Schwinger-Dewitt
  Technique in Gauge Theories and Quantum Gravity}},  {\em Phys. Rept.} {\bf
  119} (1985) 1--74.

\bibitem{Barvinsky:1988ds}
A.~O. Barvinsky and G.~A. Vilkovisky, {\it {The Effective Action in Quantum
  Field Theory: Two Loop Approximation}},  {\em Batalin, I. A. (Ed.) et al.:
  Quantum Field Theory and Quantum Statistics} {\bf 1} (1988) 245--275.

\bibitem{Barvinsky:1994hw}
A.~O. Barvinsky, {\relax Yu}.~V. Gusev, G.~A. Vilkovisky, and V.~V. Zhytnikov,
  {\it {The Basis of nonlocal curvature invariants in quantum gravity theory.
  (Third order.)}},  {\em J. Math. Phys.} {\bf 35} (1994) 3525--3542,
  [\href{http://arxiv.org/abs/gr-qc/9404061}{{\tt gr-qc/9404061}}].

\bibitem{Barvinsky:1994cg}
A.~O. Barvinsky, {\relax Yu}.~V. Gusev, G.~A. Vilkovisky, and V.~V. Zhytnikov,
  {\it {The One loop effective action and trace anomaly in four-dimensions}},
  {\em Nucl. Phys.} {\bf B439} (1995) 561--582,
  [\href{http://arxiv.org/abs/hep-th/9404187}{{\tt hep-th/9404187}}].

\bibitem{Barvinsky:1995it}
A.~O. Barvinsky, A.~G. Mirzabekian, and V.~V. Zhytnikov, {\it {Conformal
  decomposition of the effective action and covariant curvature expansion}},
  in {\em {Quantum gravity. Proceedings, 6th Seminar, Moscow, Russia, June
  12-19, 1995}}, 1995.
\newblock \href{http://arxiv.org/abs/gr-qc/9510037}{{\tt gr-qc/9510037}}.

\bibitem{Donoghue:2015xla}
J.~F. Donoghue and B.~K. El-Menoufi, {\it {QED trace anomaly, non-local
  Lagrangians and quantum Equivalence Principle violations}},  {\em JHEP} {\bf
  05} (2015) 118, [\href{http://arxiv.org/abs/1503.06099}{{\tt
  arXiv:1503.06099}}].

\bibitem{Donoghue:2015nba}
J.~F. Donoghue and B.~K. El-Menoufi, {\it {Covariant non-local action for
  massless QED and the curvature expansion}},  {\em JHEP} {\bf 10} (2015) 044,
  [\href{http://arxiv.org/abs/1507.06321}{{\tt arXiv:1507.06321}}].

\bibitem{El-Menoufi:2015ztk}
B.~K. El-Menoufi, {\it {Inflationary magnetogenesis and non-local actions: The
  conformal anomaly}},  {\em JCAP} {\bf 1602} (2016) 055,
  [\href{http://arxiv.org/abs/1511.02876}{{\tt arXiv:1511.02876}}].

\bibitem{Bautista:2017enk}
T.~Bautista, A.~Benevides, and A.~Dabholkar, {\it {Nonlocal Quantum Effective
  Actions in Weyl-Flat Spacetimes}},
  \href{http://arxiv.org/abs/1711.00135}{{\tt arXiv:1711.00135}}.

\bibitem{Fradkin:1978yf}
E.~S. Fradkin and G.~A. Vilkovisky, {\it {Conformal Invariance and Asymptotic
  Freedom in Quantum Gravity}},  {\em Phys. Lett.} {\bf B77} (1978) 262.

\bibitem{Paneitz:2008}
S.~Paneitz, {\it {A Quartic Conformally Covariant Differential Operator for
  Arbitrary Pseudo-Riemannian Manifolds (Summary)}},  {\em SIGMA} {\bf 4}
  (2008) 36, [\href{http://arxiv.org/abs/0803.4331}{{\tt 0803.4331}}].

\bibitem{Riegert:1984kt}
R.~Riegert, {\it {A Nonlocal Action for the Trace Anomaly}},  {\em Phys.Lett.}
  {\bf B134} (1984) 56--60.

\bibitem{Demozzi:2009fu}
V.~Demozzi, V.~Mukhanov, and H.~Rubinstein, {\it {Magnetic fields from
  inflation?}},  {\em JCAP} {\bf 0908} (2009) 025,
  [\href{http://arxiv.org/abs/0907.1030}{{\tt arXiv:0907.1030}}].

\bibitem{Grishchuk:1990bj}
L.~P. Grishchuk and {\relax Yu}.~V. Sidorov, {\it {Squeezed quantum states of
  relic gravitons and primordial density fluctuations}},  {\em Phys. Rev.} {\bf
  D42} (1990) 3413--3421.

\bibitem{Green:2015fss}
D.~Green and T.~Kobayashi, {\it {Constraints on Primordial Magnetic Fields from
  Inflation}},  {\em JCAP} {\bf 1603} (2016), no.~03 010,
  [\href{http://arxiv.org/abs/1511.08793}{{\tt arXiv:1511.08793}}].

\bibitem{Maldacena:2015bha}
J.~Maldacena, {\it {A model with cosmological Bell inequalities}},  {\em
  Fortsch. Phys.} {\bf 64} (2016) 10--23,
  [\href{http://arxiv.org/abs/1508.01082}{{\tt arXiv:1508.01082}}].

\bibitem{Kobayashi:2014sga}
T.~Kobayashi, {\it {Primordial Magnetic Fields from the Post-Inflationary
  Universe}},  {\em JCAP} {\bf 1405} (2014) 040,
  [\href{http://arxiv.org/abs/1403.5168}{{\tt arXiv:1403.5168}}].

\bibitem{Akrami:2018odb}
{\bf Planck} Collaboration, Y.~Akrami et~al., {\it {Planck 2018 results. X.
  Constraints on inflation}},  \href{http://arxiv.org/abs/1807.06211}{{\tt
  arXiv:1807.06211}}.

\bibitem{Patrignani:2016xqp}
{\bf Particle Data Group} Collaboration, C.~Patrignani et~al., {\it {Review of
  Particle Physics}},  {\em Chin. Phys.} {\bf C40} (2016), no.~10 100001.

\bibitem{Dolgov:1981nw}
A.~D. Dolgov, {\it {Conformal Anomaly and the Production of Massless Particles
  by a Conformally Flat Metric}},  {\em Sov. Phys. JETP} {\bf 54} (1981)
  223--228. [Zh. Eksp. Teor. Fiz.81,417(1981)].

\bibitem{Allen:1999xw}
B.~Allen, E.~E. Flanagan, and M.~A. Papa, {\it {Is the squeezing of relic
  gravitational waves produced by inflation detectable?}},  {\em Phys. Rev.}
  {\bf D61} (2000) 024024, [\href{http://arxiv.org/abs/gr-qc/9906054}{{\tt
  gr-qc/9906054}}].

\end{thebibliography}\endgroup
\end{document}